# Finite size and intrinsic field effect on the polar-active properties of the ferroelectric-semiconductor heterostructures

A.N. Morozovska[1*], E.A. Eliseev[1,2], S.V. Svechnikov[1], V.Y. Shur[3], A.Y. Borisevich[4], P. Maksymovych[4], and S.V. Kalinin[4†],

[1]*Institute of Semiconductor Physics, National Academy of Sciences of Ukraine, 03028, Kiev, Ukraine*

[2]*Institute for Problems of Materials Science, National Academy of Sciences of Ukraine, 03142, Kiev, Ukraine*

[3]*Institute of Physics and Applied Mathematics, Ural State University, 620083, Ekaterinburg, Russia*

[4]*Oak Ridge National Laboratory, 37831, Oak Ridge, Tennessee, USA*

**Abstract**

Using Landau-Ginzburg-Devonshire approach we calculated the equilibrium distributions of electric field, polarization and space charge in the ferroelectric-semiconductor heterostructures containing proper or incipient ferroelectric thin films. The role of the polarization gradient and intrinsic surface energy, interface dipoles and free charges on polarization dynamics are specifically explored. The intrinsic field effects, which originated at the ferroelectric-semiconductor interface, lead to the surface band bending and result into the formation of depletion space-charge layer near the semiconductor surface. During the local polarization reversal (caused by the inhomogeneous electric field induced by the nanosized tip of the Scanning Probe Microscope (SPM) probe) the thickness and charge of the interface layer drastically changes, it particular the sign of the screening carriers is determined by the polarization direction. Obtained analytical solutions could be extended to analyze polarization-mediated electronic transport.

---

[*] morozo@i.com.ua
[†] sergei2@ornl.gov

**1. Introduction**

Polar discontinuity at the interfaces induced either by translational symmetry breaking of a ferroelectric material or ionic charge mismatch between component can produce intriguing modification of the of the interfacial electronic states and polarization of the adjacent materials [1, 2, 3, 4]. The representative interfacial phenomena arising from the interplay between polarity and electronic structure include two-dimensional electron gases at the interface of band ($LaAlO_3/SrTiO_3$) [2] or band and Mott insulators [5] and polarization-controlled electron tunneling across ferroelectric-semiconductor or bad-metal interfaces ($PbTiO_3/(La,Sr)MnO_3$, $BaTiO_3/SrRuO_3$). [6, 7, 8, 9] These novel physical phenomena emerging in oxide materials at the nanometer scale hold strong potential for novel devices. Correspondingly, the theoretical insight into the epitaxial interfaces of normal and incipient ferroelectrics with bad metals and semiconductors and interplay between atomistic phenomena at interfaces and mesoscopic potential and field distributions is acutely needed.

As an illustrative example, the notorious problem of the Schottky barrier in ferroelectric films is still widely debated, with the key question of whether sub-100 nm films are fully depleted, or that the width of the depletion regions is in the sub-10 nm range due to the overall high density $>10^{20}$ cm$^{-3}$ of shallow and deep donor and/or acceptor levels in the film, and particularly in the interfacial regions. [10] More importantly, only several previous works, such as the concept of a ferroelectric Schottky diode [11] and the dielectric non-linearities in the epitaxial PZT films [12, 13], have emphasized the effect of space-charge layers on polarization distribution and domain switching dynamics inside ferroelectric films.

In the current paper we present analytical calculations of the polar-active properties (including local polarization reversal) in the proper and incipient ferroelectric-dielectric thin films within Landau-Ginzburg-Devonshire phenomenological approach, with a special attention to the polarization gradient and intrinsic surface energy, interface dipoles and free charges. We analyzed the influence of finite size effect on the intrinsic electric field and polar-active properties of the ferroelectric-semiconductor heterostructures. Although the stability of the spontaneous polarization in the system ferroelectric film /insulator/semiconductor was previously studied within LGD approach [14], the band structure of the system, interface charges and dipoles, the polarization gradient and intrinsic surface energy of ferroelectric film were previously ignored. Hence, this work provides a framework to link the mesoscopic LGD-semiconductor theory to the first–principle calculations that can reveal the electrostatic details of the interface structure.

The paper is organized as follows. After stating the problem in Section 2, analytical solutions for polarization, electric potential, field and space charge distributions in the model heterostructure are presented in the Section 3.1. The results of the stable ground state calculations for $SrTiO_3$/(La,Sr)$MnO_3$ (STO/LSMO) and $BiFeO_3$/(La,Sr)$MnO_3$ (BFO/LSMO) heterostructures are presented Section 3.2. Metastable states are considered in Section 3.3. The effect of the incomplete screening and interface charge on the local polarization reversal and domain formation caused by the electric field of the SPM tip is studied in Section 4. The tunneling current density is estimated, providing an analytical approach to quantify recent experimental measurements of polarization-controlled tunneling [6,7,8,9].

## 2. The problem statement

Here we consider an asymmetric heterostructure consisting of a narrow-gap (or metallic) semiconductor and a thin ferroelectric film of thickness $L$. We will consider the two cases of the proper and incipient ferroelectric films, both either a wide-gap semiconductor or a dielectric (i.e. semiconducting properties of ferroelectric are neglected). In the initial state, the external bias is absent and the free ferroelectric surface $z = -L$ is completely screened by the ambient sluggish charges [Fig. 1a]. Then inhomogeneous external bias $U(x,y)$ is applied to the tip electrode. The bias increase may cause local polarization reversal below the tip that finally results into cylindrical domain formation in thin ferroelectric film [see the final state in Fig. 1c]. Note that while we consider the tip-induced polarization switching, the obtained solutions are applicable for capacitor geometry in the limit of uniform potential.

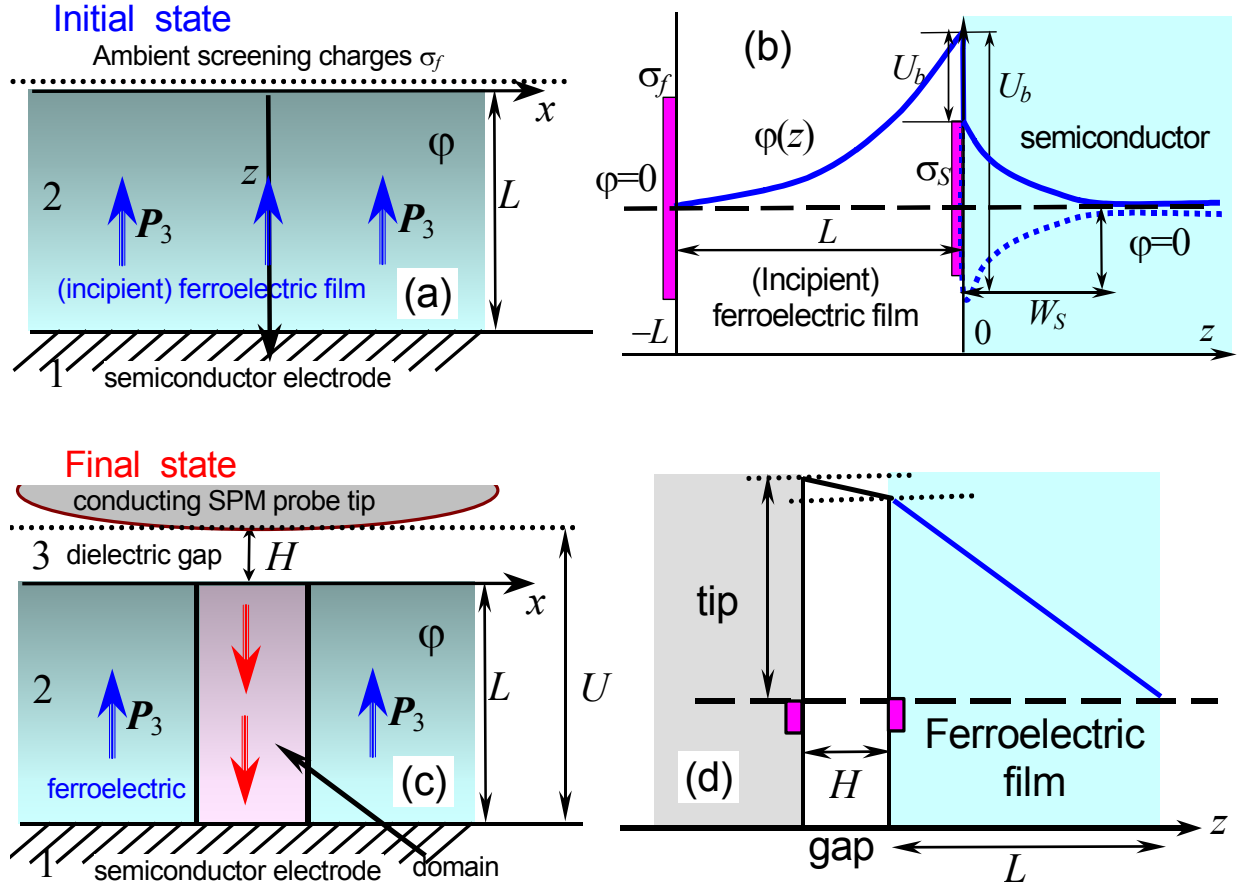

**Fig. 1.** (Color online). (a) The initial state of the considered heterostructure: semiconductor/(incipient or proper) ferroelectric-dielectric film of thickness $L$. $P_3$ is the ferroelectric polarization. (b) Sketch of the electrostatic potential distribution: $\varphi$ is the electrostatic potential, $U_b$ is the contact potential difference on ferroelectric-semiconductor interface, $W_S$ is the electric field penetration depth into the semiconductor, i.e. the thickness of the space charge depletion layer (shown by dashed and solid lines for different charge signs). (c) The final state of the ferroelectric film is the local polarization reversal caused by the biased SPM probe. (d) During the tip-induced polarization reversal an ultra-thin dielectric gap $H$ between the tip electrode and the ferroelectric surface could exist. Squares denote the screening surface charges $\sigma_f$ and the opposite charge accommodated at the tip surface.

The contact potential difference $U_b$ at the interface $z = 0$ originates from the band mismatch between ferroelectric and semiconductor, the interface bonding effect and interfacial polarity [Fig. 1b]. The band bending (or *intrinsic field effect*) in the semiconductor leads to the depletion (or accumulation) charged layers of thickness $W_S$ with space charge $\rho_S$.

The screening interface charge $\sigma_S$ could originate at $z = 0$ self-consistently in the case of the bad screening from the semiconductor side (i.e. for thick depletion layer created by the minor-type carriers). The non-ideal screening that causes the strong depolarization field controls the self-consistent mechanism. The field decreases the polarization inside the ferroelectric film and increases the free energy of the system, since depolarization field energy is always positive. As a result, the strong field effect may lead to the bend bending at $z = 0$ and appearance of charge states at the interface. The interface charges $\sigma_S$ of appropriate sign provide effective screening of the spontaneous polarization, make it more homogeneous and thus decrease the depolarization field, which in turn self-consistently decreases the system free energy. The density of the interface charge $\sigma_S$ depends on energetic position of chemical potential $\mu$ at the surface that modifies the Shottky barrier. The potential $\mu$ is manly determined by the interface layers with the energy density $N_S$ (per unit energy) of quasi-continuous surface states and Fermi level $E_F$ at the neutral surface [15, 16, 17].

We assume that in the initial state the sluggish surface charges $\sigma_f$ completely screen the electric displacement outside the film, i.e. $\sigma_f(x,y,-L) = -D_3(x,y,-L)$, $\varphi(x,y,-L) = 0$ and $D_3(z < -L) = 0$. This behavior is analyzed in Section 3. In contrast, recharging of the surface charges $\sigma_f$ should appear during the polarization reversal. The ultra-thin dielectric gap of thickness $H$ models the resistive properties of the sluggish surface charges $\sigma_f$, contamination or dead layer. Corresponding free image charges $-\sigma_f$ are accommodated at the conducting SPM tip surface. Without loss of generality one can assume that the equilibrium domain structure is almost cylindrical for the case of complete local polarization reversal in thin ferroelectric film. The assumptions significantly simplify the problem considered in the Section 4, and allows developing the analytical description for the domain formation.

Hereinafter we assume that the time of external field changing is small enough for the validity of the quasi-static approximation $rot\,\mathbf{E} \approx 0$. Maxwell's equations for the quasi-static electric field $\mathbf{E} = -\nabla\varphi$ and displacement $\mathbf{D}$ inside semiconductors have the form:

$$div\,\mathbf{D} = div(\varepsilon_0 \mathbf{E} + \mathbf{P}) = \rho(\varphi). \tag{1}$$

Here the electrostatic potential $\varphi(x,y,z)$ is determined by external bias as well as by contact and surface effects. The potential determines the free carrier density $\rho(\varphi)$ determined by the concentration of holes in the valence band, electrons in the conduction band, and acceptors and donors at their respective levels in the band gap [16].

The ferroelectric film that occupies the region $-L < z < 0$ is transversally isotropic, i.e. permittivity $\varepsilon_{11} = \varepsilon_{22}$ at zero electric field. We further assume that the dependence of in-plane polarization components on $E_{1,2}$ can be linearized as $P_{1,2} \approx \varepsilon_0 (\varepsilon_{11} - 1) E_{1,2}$ ($\varepsilon_0$ is the universal dielectric constant), while the polarization component $P_3$ nonlinearly depends on external field. Thus corresponding polarization vector acquires the form:
$\mathbf{P}(\mathbf{r}) = \left( \varepsilon_0 (\varepsilon_{11} - 1) E_1, \ \varepsilon_0 (\varepsilon_{11} - 1) E_2, \ P_3(\mathbf{E}, \mathbf{r}) + \varepsilon_0 (\varepsilon_{33}^b - 1) E_3 \right)$ [18].

Within the framework of the Landau-Ginzburg-Devonshire (LGD) theory, quasi-equilibrium polarization distribution $P_3(x,y,z)$ in the ferroelectric film with the spatial dispersion should be found from the Euler-Lagrange boundary problem:

$$\begin{cases} \alpha P_3 + \beta P_3^3 - g \left( \Delta_\perp + \dfrac{\partial^2}{\partial z^2} \right) P_3 = -\dfrac{\partial \varphi}{\partial z}, \\ \left( P_3 + \lambda_1 \dfrac{\partial P_3}{\partial z} \right) \bigg|_{z=0} = -P_b, \quad \left( P_3 - \lambda_2 \dfrac{\partial P_3}{\partial z} \right) \bigg|_{z=-L} = 0. \end{cases} \quad (2)$$

Hereinafter we introduced a transverse Laplace operator $\Delta_\perp = \dfrac{\partial^2}{\partial x^2} + \dfrac{\partial^2}{\partial y^2}$.

The temperature-dependent coefficient $\alpha$ is positive for incipient ferroelectric and proper ferroelectrics in paraelectric phase, while $\alpha < 0$ for proper ferroelectrics in ferroelectric phase, $\beta > 0$ for the second order ferroelectrics considered hereinafter, gradient coefficient $g > 0$. Extrapolation lengths $\lambda_{1,2}$ originate from the surface energy coefficients in the LGD-free energy.

Inhomogeneity $P_b$ describes the effect of the interface polarization stemming from the interface bonding effect and associated interface dipole [19, 20]. More generally, the translation symmetry breaking inevitably present in the vicinity of the any interface will give rise to inhomogeneity in the boundary conditions (2) [21, 22].

Eqs. (1)-(2) yield the coupled system:

$$\left(\frac{\partial^2 \varphi}{\partial z^2} + \Delta_\perp \varphi\right) = 0, \quad -H-L < z < -L,$$

$$\varepsilon_{33}^b \frac{\partial^2 \varphi}{\partial z^2} + \varepsilon_{11}\Delta_\perp \varphi = \frac{1}{\varepsilon_0}\left(\frac{\partial P_3}{\partial z} - \rho_f(\varphi)\right), \quad -L < z < 0, \quad (3)$$

$$\varepsilon_0 \varepsilon_S \left(\frac{\partial^2 \varphi}{\partial z^2} + \Delta_\perp \varphi\right) = -\rho_S(\varphi), \quad z > 0.$$

The background dielectric permittivity of (incipient) ferroelectric $\varepsilon_{33}^b$ (typically $\varepsilon_{33} \gg \varepsilon_{33}^b$); $\varepsilon_S$ is the semiconductor (bare) lattice permittivity.

Eqs.(3) are supplemented with the boundary conditions at $z = -L-H$, $z = -L$, $z = 0$ and $z = +\infty$, namely

$$\varphi(x,y,-L-H) = U_e(x,y), \quad \varphi(x,y,-L+0) = \varphi(x,y,-L-0), \quad \varphi(x,y,z\to\infty) = 0, \quad (4a)$$

$$\varphi(x,y,+0) - \varphi(x,y,-0) = U_b, \quad (4b)$$

$$\varepsilon_0 \varepsilon_{33}^b \frac{\partial \varphi(x,y,-0)}{\partial z} - P_3(x,y,-0) - \varepsilon_S \varepsilon_0 \frac{\partial \varphi(x,y,+0)}{\partial z} = \sigma_S(x,y), \quad (4c)$$

$$-\varepsilon_0 \varepsilon_{33}^b \frac{\partial \varphi(x,y,-L+0)}{\partial z} + P_3(x,y,-L+0) + \varepsilon_{33}^g \varepsilon_0 \frac{\partial \varphi(x,y,-L-0)}{\partial z} = \sigma_f(x,y). \quad (4d)$$

Where $U_b$ is the contact potential difference at the dielectric-semiconductor interface. $\varepsilon_{33}^g$ is the dielectric constant of the dielectric gap between the tip and ferroelectric surface. The potential distribution $U_e(x,y)$ produced by the SPM tip is assumed to be almost constant in the surface spatial region much larger then the film thickness.

## 3. Solution for polarization, electric potential, field and space charge distributions
### 3.1. Analytical solutions for the one-dimensional case

Here we calculate the potential and polarization distribution in the initial ground state in the one-dimensional case $U_e$ = const that corresponds to the plain electrodes. The case is realized in paraelectric or incipient ferroelectric film as well as in the monodomain state of the proper ferroelectric thin film.

The space charge density inside the doped p-type (or n-type) semi-infinite semiconductor has the form

$$\rho_S(\varphi) = q\big(p(\varphi) + N_d^+(\varphi) - n(\varphi) - N_a^-(\varphi)\big),$$

$$p(\varphi) = N_p^0 \cdot F\left(\frac{E_F - E_V + q\varphi}{k_B T}\right), \quad N_d^+(\varphi) = N_d \cdot F\left(\frac{E_F - E_d + q\varphi}{k_B T}\right), \quad (5)$$

$$n(\varphi) = N_e^0 \cdot F\left(\frac{E_C - E_F - q\varphi}{k_B T}\right), \quad N_a^-(\varphi) = N_a \cdot F\left(\frac{E_a - E_F - q\varphi}{k_B T}\right).$$

Where $F(\theta) = (\exp(\theta) + 1)^{-1}$ is the Fermi-Dirac distribution function, $q$ is the absolute value of the carrier elementary charge. $E_F$, $E_V$, $E_C$, $E_d$ and $E_a$ are the energies of Fermi level, valence band, conductance band, donor and acceptor levels in the quasi-neutral region of the semiconductor correspondingly. Since $\rho_S(\varphi) \to 0$ in the quasi-neutral region of the semiconductor, where $\varphi \to 0$, the identity $p(0) + N_d^+(0) - n(0) - N_a^-(0) = 0$ should be valid. The identity along with typical assumption $N_d^+ \approx const$, $N_a^- \approx const$ and Boltzman approximation for electrons $E_C - E_F - q\varphi \gg k_B T$ or holes $E_F - E_V + q\varphi \gg k_B T$ lead to expressions $\rho_S \approx qp_S^0\left(\exp\left(-\frac{q\varphi}{k_B T}\right) - 1\right)$ or $\rho_S \approx -qn_S^0\left(\exp\left(\frac{q\varphi}{k_B T}\right) - 1\right)$ correspondingly, where $p_S^0$ and $n_S^0$ are equilibrium concentrations of holes and electrons in the quasi-neutral region of the semiconductor [23].

Then in depletion layer (or abrupt junction) approximation the space charge density near the interface of the strongly doped p-type (or n-type) semi-infinite semiconductor has the form

$$\rho_S(\varphi) \approx \begin{cases} \rho_S^0, & 0 < |z| < W_S \\ 0, & |z| > W_S \end{cases} \quad (5b)$$

Hereinafter the choice of the charge density $\rho_S^0 = qp_S^0$ and depth $W_S = W_{Sp}$ (or $\rho_S^0 = -qn_S^0$ and $W_S = W_{Sn}$) is determined by the sign of potential (i.e. by the sign of charge in depletion layer). More rigorously, for the definite type of carriers the thicknesses of the depletion layers $W_S$ (i.e. the field penetration depths) should be determined self-consistently from the exact solution of the system (3)-(5).

The ferroelectric film is regarded as a wide-gap proper semiconductor or almost dielectric, so its space-charge density is negligibly small: $\rho_f(\varphi) \approx 0$ at $-L < z < 0$. The nonzero 1D-solution of Eqs. (3) with respect to the boundary conditions (4a) and (4c-e) is

$$\varphi(z) \approx \begin{cases} -\dfrac{\rho_S^0}{2\varepsilon_0\varepsilon_S}(W_S - z)^2\,\theta(W_S - z), & z > 0, \\ \displaystyle\int_{-L}^{z}\dfrac{P_3(\tilde{z})}{\varepsilon_0\varepsilon_{33}^b}d\tilde{z} - (L+z)\left(\dfrac{\rho_S^0 W_S - \sigma_S}{\varepsilon_0\varepsilon_{33}^b}\right) + U_e + \dfrac{H}{\varepsilon_0\varepsilon_{33}^g}(\sigma_S + \sigma_f - \rho_S^0 W_S), & -L \leq z < 0, \\ U_e + \dfrac{H + L + z}{\varepsilon_0\varepsilon_{33}^g}(\sigma_S + \sigma_f - \rho_S^0 W_S), & -L - H \leq z < -L. \end{cases} \quad (6)$$

Here $\theta(z)$ is the step-function. So the semiconductor potential and space charge are distributed in the layer $0 < z < W_S$ and zero outside. Approximate expressions (6) for the potential $\varphi$ correspond to parabolic approximation valid in the depletion/accumulation limit, at that $W_{Sn} \ll W_{Sp}$ or $W_{Sn} \gg W_{Sp}$ depending of the main carriers $n$ or $p$-type. Note, that in the accumulation regime the interface charge $\sigma_S$ is localized in the thin depletion layer of several nm (for oxide electrodes) thickness $W_S$ that is occupied by the main-type carriers. The opposite case of charged layers created by the minor-type carriers, which can appear during the polarization reversal, could lead the strong band bending and accommodation of $\sigma_S$.

From Eq.(6) we derived the electric field $E_3$ and electrical displacement $D_3$ distributions in the parabolic approximation:

$$E_3(z) \approx -\dfrac{P_3(z)}{\varepsilon_0\varepsilon_{33}^b} + \dfrac{\rho_S^0 W_S - \sigma_S}{\varepsilon_0\varepsilon_{33}^b}, \quad D_3(z) = \rho_S^0 W_S - \sigma_S, \quad \text{at} \quad -L < z < 0,$$

$$E_3(z) \approx \dfrac{\rho_S^0}{\varepsilon_0\varepsilon_S}(z - W_S)\theta(W_S - z), \quad D_3(z) = \rho_S^0(z - W_S)\theta(W_S - z), \quad \text{at} \quad z > 0. \quad (7)$$

Note, that the free screening charge $-\sigma_f = D_3(-L)$ should also originate at another interface $z = -L$. Thus the condition for electroneutrality of the whole system is $-\sigma_S + \rho_S^0 W_S + \sigma_f = 0$.

Allowing for Eqs.(6)-(7) polarization distribution $P_3(z)$ was found from the Euler-Lagrange boundary problem (2) as described in Appendix A. The polarization distribution acquires the form:

$$P_3(z) = \dfrac{\varepsilon_S - 1}{\varepsilon_S}\rho_S^0(z - W_S)\cdot\theta(W_S - z), \quad z \geq 0, \quad (8a)$$

$$P_3(z) = \dfrac{2\varepsilon_0\varepsilon_{33}^b\beta\langle P_3\rangle^3 + \rho_S^0 W_S - \sigma_S}{\varepsilon_0\varepsilon_{33}^b(\alpha + 3\beta\langle P_3\rangle^2) + 1} f(z,L) - P_b \cdot b(z,L), \quad -L < z \leq 0. \quad (8b)$$

Parameter $\langle P_3 \rangle$ in Eq.(8b) is the polarization averaged over the film depth: $\langle P_3 \rangle \equiv \frac{1}{L}\int_{-L}^{0} P_3(\tilde{z})d\tilde{z}$.

Its spatial distribution is governed by the functions $f$ and $b$:

$$f(z,L) = 1 - \xi \frac{\lambda_2 \cosh((L+z)/\xi) + \lambda_1 \cosh(z/\xi) + \xi(\sinh((L+z)/\xi) - \sinh(z/\xi))}{(\xi^2 + \lambda_1\lambda_2)\sinh(L/\xi) + \xi(\lambda_1 + \lambda_2)\cosh(L/\xi)}, \quad (9a)$$

$$b(z,L) = \frac{\xi\lambda_2 \cosh((L+z)/\xi) + \xi^2 \sinh((L+z)/\xi)}{(\xi^2 + \lambda_1\lambda_2)\sinh(L/\xi) + \xi(\lambda_1 + \lambda_2)\cosh(L/\xi)}. \quad (9b)$$

Characteristic length $\xi \approx \sqrt{\varepsilon_0 \varepsilon_{33}^b g}$.

The average polarization $\langle P_3 \rangle$ and depths $W_{Sn,p}$ should be determined self-consistently from the spatial averaging of Eq.(8b) and the boundary conditions (4b). After elementary transformations we obtained the system of two coupled algebraic equations:

$$\begin{cases} \langle P_3 \rangle = \rho_S^0 W_S - \sigma_S + \frac{\varepsilon_{33}^b \rho_S^0}{2L\varepsilon_S}W_S^2 - \frac{\varepsilon_0\varepsilon_{33}^b}{L}\left(U_b + U_e + \frac{H}{\varepsilon_0\varepsilon_{33}^g}(\sigma_S + \sigma_f - \rho_S^0 W_S)\right), \\ \left(\alpha + \frac{1}{\varepsilon_0\varepsilon_{33}^b}\right)\langle P_3 \rangle + \beta\langle P_3 \rangle^3(3 - 2\langle f \rangle) = \left(\frac{\rho_S^0 W_S - \sigma_S}{\varepsilon_0\varepsilon_{33}^b}\right)\langle f \rangle - \frac{P_b}{\varepsilon_0\varepsilon_{33}^b}\langle b \rangle. \end{cases} \quad (10)$$

Where the average values $\langle f \rangle \approx 1 - \frac{\xi^2(2\xi + \lambda_1 + \lambda_2)}{L(\xi(\lambda_1 + \lambda_2) + \xi^2 + \lambda_1\lambda_2)}$ and $\langle b \rangle \approx \frac{\xi^2}{L(\xi + \lambda_1)}$.

In particular case of narrow-gap metallic semiconductor and thick enough ferroelectric film the strong inequality $W_S \ll L$ is valid. This leads to the linear approximation $\langle P_3 \rangle \approx (\rho_S^0 W_S - \sigma_S)\left(1 + \frac{\varepsilon_{33}^b H}{\varepsilon_{33}^g L}\right) - \frac{\varepsilon_0\varepsilon_{33}^b}{L}\left(U_b + U_e + \frac{H\sigma_f}{\varepsilon_0\varepsilon_{33}^g}\right)$ and the cubic equation for the average polarization:

$$\left(\alpha + \frac{1}{\varepsilon_0\varepsilon_{33}^b}\left(1 - \frac{\varepsilon_{33}^g L\langle f \rangle}{\varepsilon_{33}^g L + \varepsilon_{33}^b H}\right)\right)\langle P_3 \rangle + \beta\langle P_3 \rangle^3(3 - 2\langle f \rangle) = E_b^f(L,H) + E_e^f(L,H). \quad (11a)$$

Where the built-in electric field $E_b^f(L,H)$ and external field $E_e^f(L,H)$ are introduced as:

$$E_b^f(L,H) = \frac{\varepsilon_{33}^g \langle f \rangle}{\varepsilon_{33}^g L + \varepsilon_{33}^b H}\left(U_b + \frac{H\sigma_f}{\varepsilon_0\varepsilon_{33}^g}\right) - \frac{P_b}{\varepsilon_0\varepsilon_{33}^b}\langle b \rangle, \quad (11b)$$

$$E_e^f(L,H) = \frac{\varepsilon_{33}^g \langle f \rangle U_e}{\varepsilon_{33}^g L + \varepsilon_{33}^b H}. \tag{11c}$$

Note that Eq.(11a) this is equivalent to the equation for ferroelectric polarization hysteresis loop in the uniform electric field, while the built-in field $E_b^f(L,H)$ determines the horizontal imprint of the loop. Thus for the case considered by Eqs.(11) the symmetric intrinsic coercive fields $E_c^b = \pm \frac{2}{3\sqrt{3}}\sqrt{-\frac{\alpha^3}{\beta}}$ of a bulk material [24] become asymmetric and has the form.

$$E_c^\pm(L,H,T) = -E_b^f(L,H) \pm \frac{2}{3\sqrt{3}}\sqrt{-\frac{1}{\beta(3-2\langle f \rangle)}\left(\alpha(T) + \frac{1}{\varepsilon_0 \varepsilon_{33}^b}\left(1 - \frac{\varepsilon_{33}^g L \langle f \rangle}{\varepsilon_{33}^g L + \varepsilon_{33}^b H}\right)\right)^3}. \tag{12}$$

Renormalization of the coefficient α in Eq.(11), i.e. the term $\frac{1}{\varepsilon_0 \varepsilon_{33}^b}\left(1 - \frac{\varepsilon_{33}^g L \langle f \rangle}{\varepsilon_{33}^g L + \varepsilon_{33}^b H}\right)$, quantitatively reflects the "extrinsic contribution" (factor $0 < \frac{\varepsilon_{33}^g L}{\varepsilon_{33}^g L + \varepsilon_{33}^b H} < 1$) originated from the depolarization field produced by the finite gap $H$ and the "intrinsic contribution" (factor $0 < \langle f \rangle \leq 1$) originated from the finite extrapolation lengths $\lambda_i < \infty$ and intrinsic polarization gradient ($\langle f \rangle = 1$ for either g=0 or both $\lambda_i = \infty$). Thus Eq.(11) allows rigorous estimations of both extrinsic and intrinsic contributions into the renormalization of the transition temperature into paraelectric phase (compare the result with Ref.[25]).

For proper ferroelectrics the coefficient $\alpha(T) = \alpha_T(T - T_c^*)$, where $T$ is the absolute temperature and $T_c^*$ is the Curie temperature possibly renormalized by misfit strain originated from the film and semiconductor substrate lattice mismatch. For instance, $T_c^* = T_c + \frac{2u_m}{\alpha_T}\frac{q_{12}c_{11} - c_{12}q_{11}}{c_{11}}$, where $u_m$ is a misfit strain, the stiffness tensor $c_{ijkl}$ is positively defined, $q_{ijkl}$ stands for the electrostriction stress tensor.

Hence, the critical thickness (as well as the critical temperature) of the size-induced transition of the ferroelectric film into paraelectric phase can be found from the condition $\alpha + \frac{1}{\varepsilon_0 \varepsilon_{33}^b}\left(1 - \frac{\varepsilon_{33}^g L \langle f \rangle}{\varepsilon_{33}^g L + \varepsilon_{33}^b H}\right) = 0$ as:

$$L_{cr}(T) \approx -\frac{1}{\varepsilon_0 \alpha(T)} \left( \frac{\xi^2(2\xi + \lambda_1 + \lambda_2)}{\varepsilon_{33}^b(\xi(\lambda_1 + \lambda_2) + \xi^2 + \lambda_1\lambda_2)} + \frac{H}{\varepsilon_{33}^g} \right), \qquad (13)$$

$$T_{cr}(L) \approx T_c^* - \frac{1}{\alpha_T \varepsilon_0 \varepsilon_{33}^b} \left( 1 - \frac{\varepsilon_{33}^g L}{\varepsilon_{33}^g L + \varepsilon_{33}^b H} \left( 1 - \frac{\xi^2(2\xi + \lambda_1 + \lambda_2)}{L(\xi(\lambda_1 + \lambda_2) + \xi^2 + \lambda_1\lambda_2)} \right) \right). \qquad (14)$$

For unstrained incipient ferroelectric film the coefficient α(T) is positive up to zero temperatures and typically is given by Barret's formula, thus the critical temperature (as well as the critical thickness) does not exist, since the film remained paraelectric up to zero Kelvin. However for the strained film it may become positive, indicative of the transition to the ferroelectric state [26].

For particular case $H = 0$ (gap is absent) the build-in field $E_b^f$ is inversely proportional to the film thickness, $E_b^f(L,0) \approx B/L$, while its value depends on the built-in polarization $P_b$, surface charge $\sigma_f$, and contact potential difference $U_b$ [see Eq.(11b) and the dashed almost straight line in Fig. 2a]. The build-in field $E_b^f$ leads to the vertical asymmetry and horizontal imprint of the polarization hysteresis loops in ferroelectric films of thickness more than critical $L > L_{cr}(T)$ [see regions 2 and 3 in Fig.2a and Fig. 2b]. Field-induced polarized state appears at film thickness less than the critical one $L < L_{cr}(T)$ [see regions 5 and 6 in Fig.2a and Fig. 2c].

For typical ferroelectric material parameters approximate expressions for the right and left coercive biases are $E_c^\pm(L,0,T) \approx -(B/L) \pm E_c^b(T)\sqrt{(1 - L_{cr}(T)/L)^3}$ [see Eq.(12)]. Thus solid curves in Fig. 2a look like an asymmetric "bird beak" with the tip $E_c^\pm(L = L_{cr}) \approx -(B/L_{cr})$ and asymptotes $E_c^\pm(L \gg L_{cr}) \to \pm E_c^b(T)$. The built-in field sign and thickness dependence determine the beak «up» or «down» asymmetry and shape correspondingly.

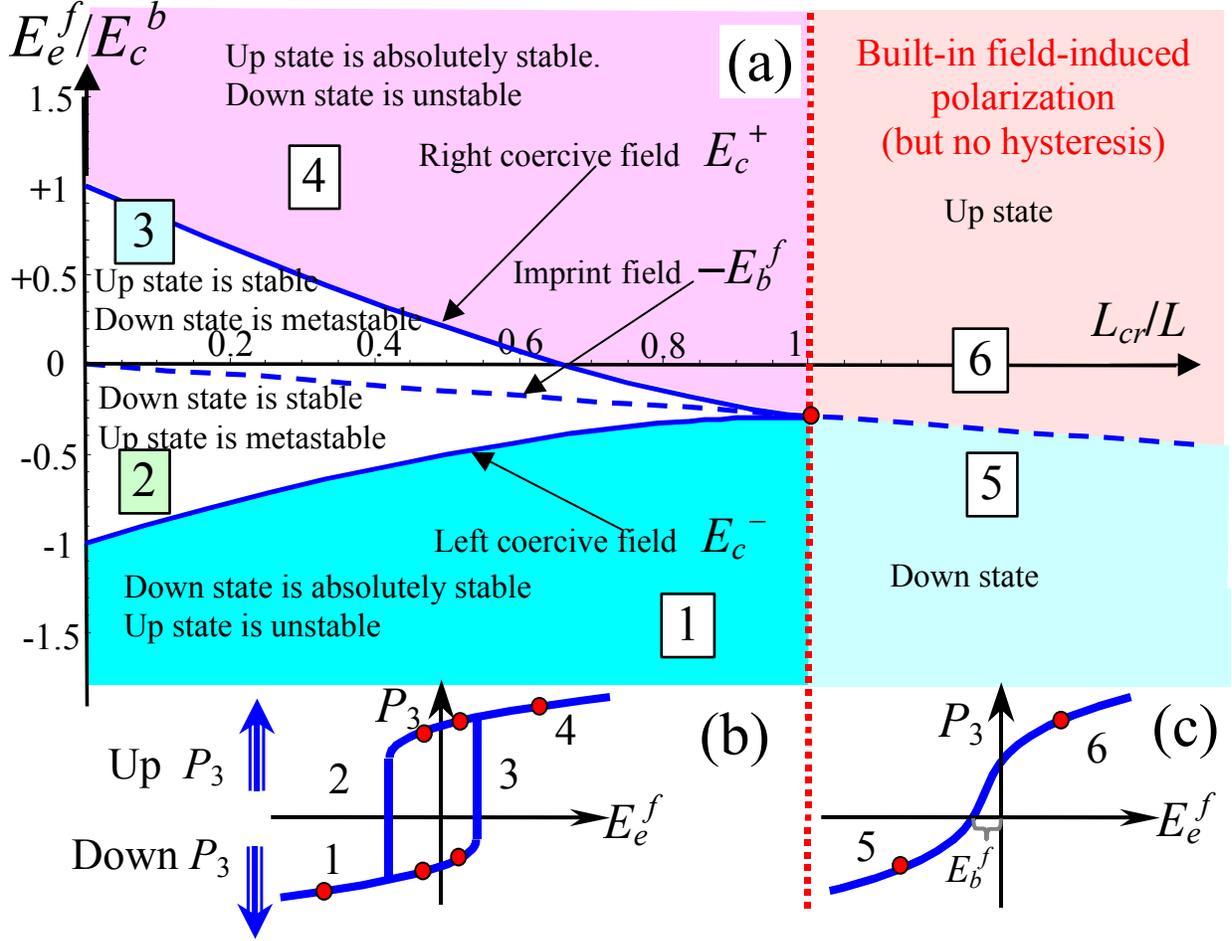

**Fig.2.** (Color online). (a) Diagram in coordinates "field – inverse thickness", $\left\{\dfrac{E_e^f}{E_c^b}, \dfrac{L_{cr}}{L}\right\}$, that shows the stability of the "up" ($\langle P_3 \rangle > 0$) and "down" ($\langle P_3 \rangle < 0$) polarization states in a ferroelectric film. Dashed curve is the thickness dependence of the built-in field $-E_b^f(L)$ calculated from Eq.(11b); solid curves are thickness dependences of right and left coercive fields $E_c^\pm(L)$ calculated from Eq.(11b) and (12) at $H=0$, fixed temperature and extrapolation lengths. $E_c^b$ is the absolute value of the bulk coercive field, $L_{cr}$ is the film critical thickness given by Eq.(13). (b-c) Hysteresis loops schematics in a ferroelectric phase (b) and field-induced polarized state (c).

Note, that there may be two possible solutions corresponding to the polarization up and down. Region 4 in Fig.2a corresponds to the stable "up" polarization states ($\langle P_3 \rangle > 0$), metastable states are absent here. The situation is vise versa in the region 1. Region 6 corresponds to the "up" polarization. The situation is vise versa in the region 5. Hysteresis loops are absent in the thickness regions 5 and 6, since at thicknesses $L < L_{cr}$ the loops never exist, here renormalized coefficient α becomes positive and coercive bias given by Eq.(12) is complex value.

Bistability of polarization states exists only in the regions 2 and 3. "Up" states are absolutely stable in the region 3, while the "down" state is metastable here. The situation is vise versa in the region 2. The depletion lengths $W_{Sn,p}$ are at least several times different for to the "up" and "down" polarization states.

In the next sections we will show how the presence of build-in field $E_b^f$ the smear the size-induced phase transition and induces polarization in the incipient ferroelectric films.

### *3.2. Calculations of the stable ground state for typical heterostructures (1D-case, $U_e = 0$)*

Here we calculate the potential and polarization distribution in the **absolutely stable** ground state, metastable states will be discussed in the next section. First, we derive the field structure for the case when the external bias is absent ($U_e = 0$). This one-dimensional case is realized in paraelectric or incipient ferroelectric film as well as in the monodomain state of the proper ferroelectric thin film. We assume that the surface charge $\sigma_f$ localized at $z = -L$ should provide the full screening of the spontaneous polarization outside the film and minimize the depolarization field energy, i.e. they acts as a perfect electrode and thus provide $\varphi(x, y, -L) = 0$, $D_3(-L) = -\sigma_f$ and $D_3(z < -L) = 0$. Thus we put $H = 0$ for the ground states calculations in Sections 3.2 and 3.3.

From relations (10) one can determine the thickness dependence of penetration depths $W_{Sp,n}$ and polarization $\langle P_3 \rangle$ for different extrapolation lengths, since the quantities $U_b$, $n_S^0$, $p_S^0$ $\varepsilon_S$ and $\varepsilon_{33}^b$ can be regarded as known material parameters. LGD-expansion coefficients α, β and the gradient coefficient $g$ are tabulated for the majority of proper and incipient ferroelectrics.

Material parameters used in the calculations of the heterostructures $SrTiO_3/(La,Sr)MnO_3$ (STO/LSMO) and $BiFeO_3/(La,Sr)MnO_3$ (BFO/LSMO) polar properties are listed in Table 1.

**Table 1.**

| Material | Permittivity | Carries concentration (cm$^{-3}$) | Band gap (eV) | LGD-expansion coefficients for ferroelectrics |
|---|---|---|---|---|
| LaSrMnO$_3$ (LSMO) half-metal | $\varepsilon_S = 30$ [27] | 1.83 10$^{22}$ [28] 1.65 10$^{21}$ [29] (La$_{0.7}$Sr$_{0.3}$MnO$_3$) | 1 $p$-type [30, 31] | Non ferroelectric |
| BiFeO$_3$ (BFO) ferroelectric | $\varepsilon_{33}^b = 30$ | wide band-gap semiconductor | 3 | $\alpha_T = 9.8 \cdot 10^5$ m/(F K) $T_c = 1103$ K $\beta = 13 \cdot 10^8$ m$^5$/(C$^2$F) $g = 10^{-8}$ m$^3$/F |
| SrTiO$_3$ (STO) incipient ferroelectric | $\varepsilon_{33}^b = 43$ [32] | dielectric | 3 (without impurities) | $\alpha_T = 1.26 \cdot 10^6$ m/(F K), $T_c = 30$ K, $\beta = 6.8 \cdot 10^9$ m$^5$/(C$^2$F) $g = 10^{-7} - 10^{-9}$ m$^3$/F |

Extrapolation lengths $\lambda_i$ and interface charge $\sigma_S$ values are not listed in the table, since they strongly depend on the interfacial states. Extrapolation length values could be extracted from the polarization distribution in ferroelectric nanosystems (films, wires, etc) obtained either experimentally [33, 34] or determined by the first principle calculations [19, 20]. Extremely small and extremely high values of extrapolation lengths describe the two limiting cases of the surface energy contribution to the total free energy. Extremely small values of $\lambda_i$ correspond to complete suppression of the polarization on the surface, while extremely large ones – to the absence of the surface energy dependence on polarization (so called natural boundary conditions). Allowing for the remark below we consider two limiting cases of the small and high values of extrapolation lengths.

The polar interface may produce a positive ionic charge leading to displacements in LSMO. Fig. 3 shows z-distributions of the electric polarization in the heterostructure "ferroelectric BiFeO$_3$ film – half-metal LaSrMnO$_3$" (BFO/LSMO) for different values of interface polarization $P_b$ and interface charge $\sigma_S$. It is seen that the polarization distribution near $z = 0$ and its asymmetry are the main effects of the interface polarization, while the interface charge creates the homogeneous electric field inside the ferroelectric film. That is why the effect of $P_b$ on the polarization distribution is much weaker than the effect of $\sigma_S$. Below we consider the case $P_b=0$.

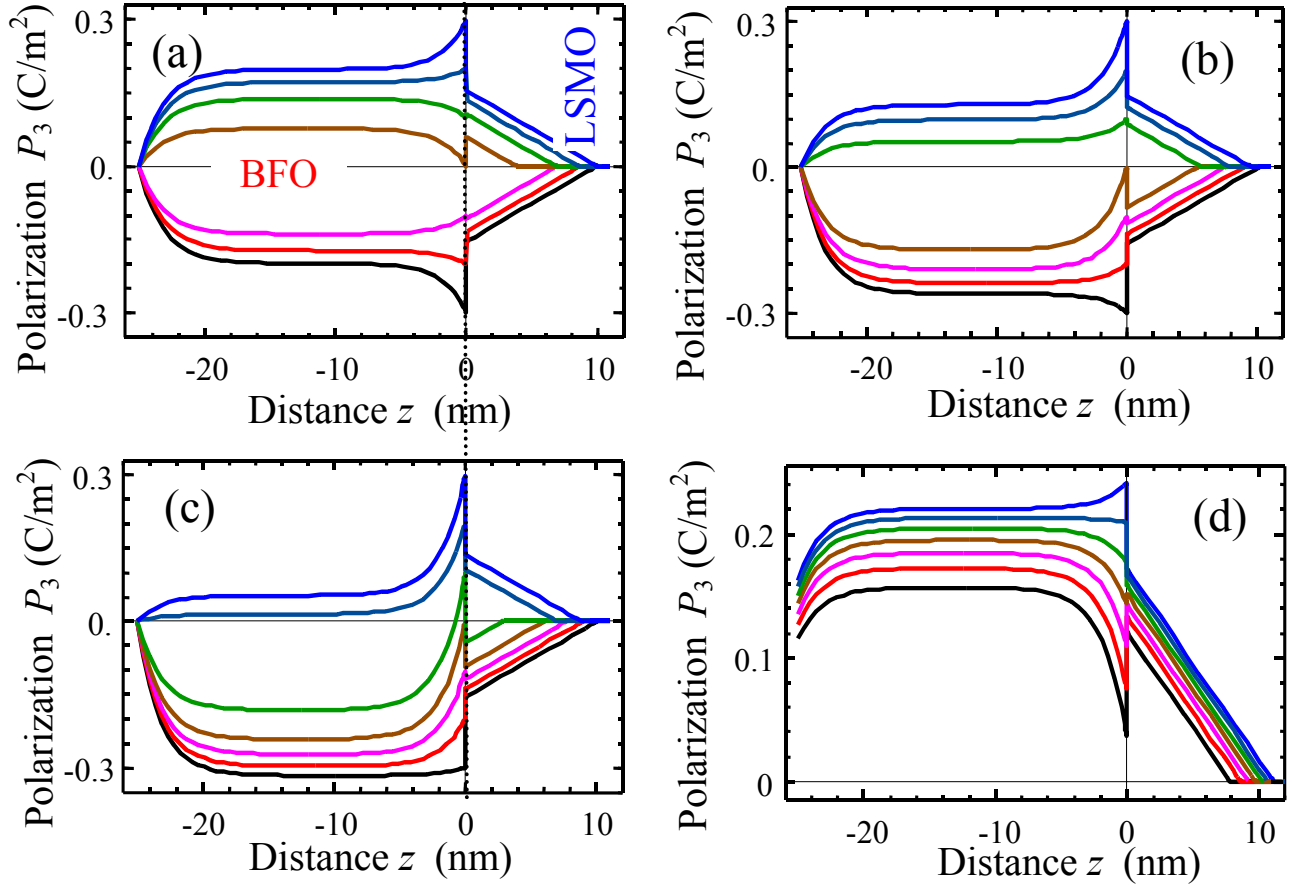

**Fig.3.** Polarization depth distribution (z in nm) for the BFO/LSMO heterostructure. Contact potential difference at $z=0$ is $U_b=0$ V, interface polarization $P_b = -0.3, -0.2, -0.1, 0, 0.1, 0.2, 0.3$ C/m$^2$ (curves from top to bottom). Carriers concentration in LSMO is $p_S^0 = 10^{26}$ m$^{-3}$, BFO thickness $L = 25$ nm and the gradient coefficient g = $10^{-8}$ m$^3$/F. (a) Extrapolation lengths $\lambda_i=0$ nm and interface charge density $\sigma_S = 0$; (b) $\sigma_S = 0.05$ C/m$^2$, $\lambda_i=0$ nm; (c) $\sigma_S = 0.1$ C/m$^2$, $\lambda_i=0$ nm; (d) $\sigma_S = 0$, $\lambda_i=5$ nm.

Fig. 4 shows z-dependence of the electric polarization, potential, field and bulk charge density in the heterostructure of BFO/LSMO for different interface polarization $P_b$, fixed extrapolation length and interface charge $\sigma_S$.

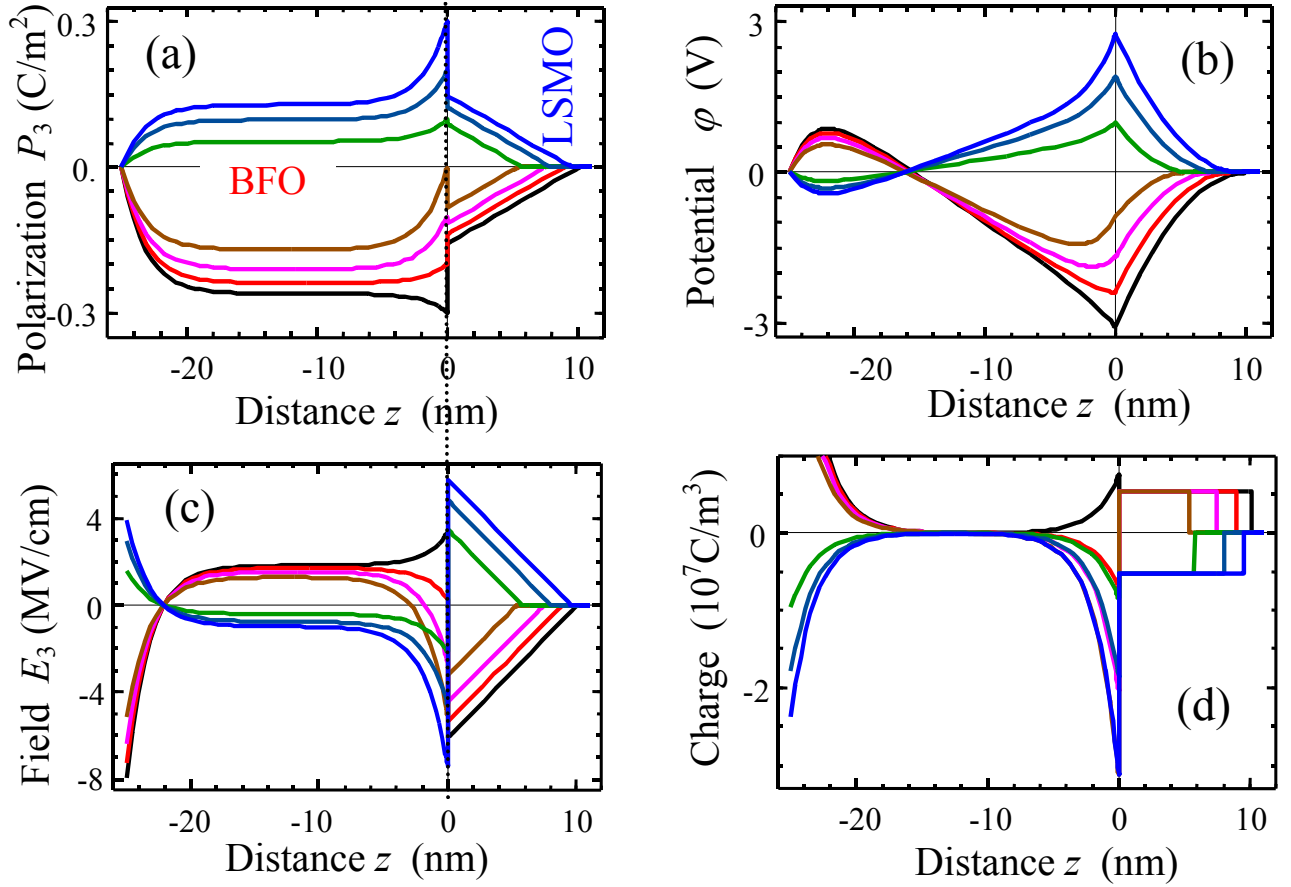

**Fig.4.** (a) Polarization, (b) potential, (c) electric field and (d) electric charge density $z$-distributions for the BFO/LSMO heterostructure. Contact potential difference at $z=0$ is $U_b=0$ V, interface charge density $\sigma_S = 0.05$ C/m$^2$ and interface polarization $P_b = -0.3, -0.2, -0.1, 0, 0.1, 0.2, 0.3$ C/m$^2$ (curves from top to bottom). Carriers concentration in LSMO is $p_S^0 = 10^{26}$ m$^{-3}$, BFO thickness $L = 25$ nm and the gradient coefficient g $= 10^{-8}$ m$^3$/F.

Figs. 5-6 show the spatial distribution of electric polarization, potential, field and bulk charge density in the heterostructure BFO/LSMO for zero interface charge $\sigma_S=0$. When generating the plots in Figs. 5 we put $\lambda_i=0$. Plots in Figs. 6 correspond to high enough $\lambda_i$ values. As anticipated the polarization distribution is almost homogeneous for the high extrapolation length. For a small extrapolation length the polarization profile is more inhomogeneous.

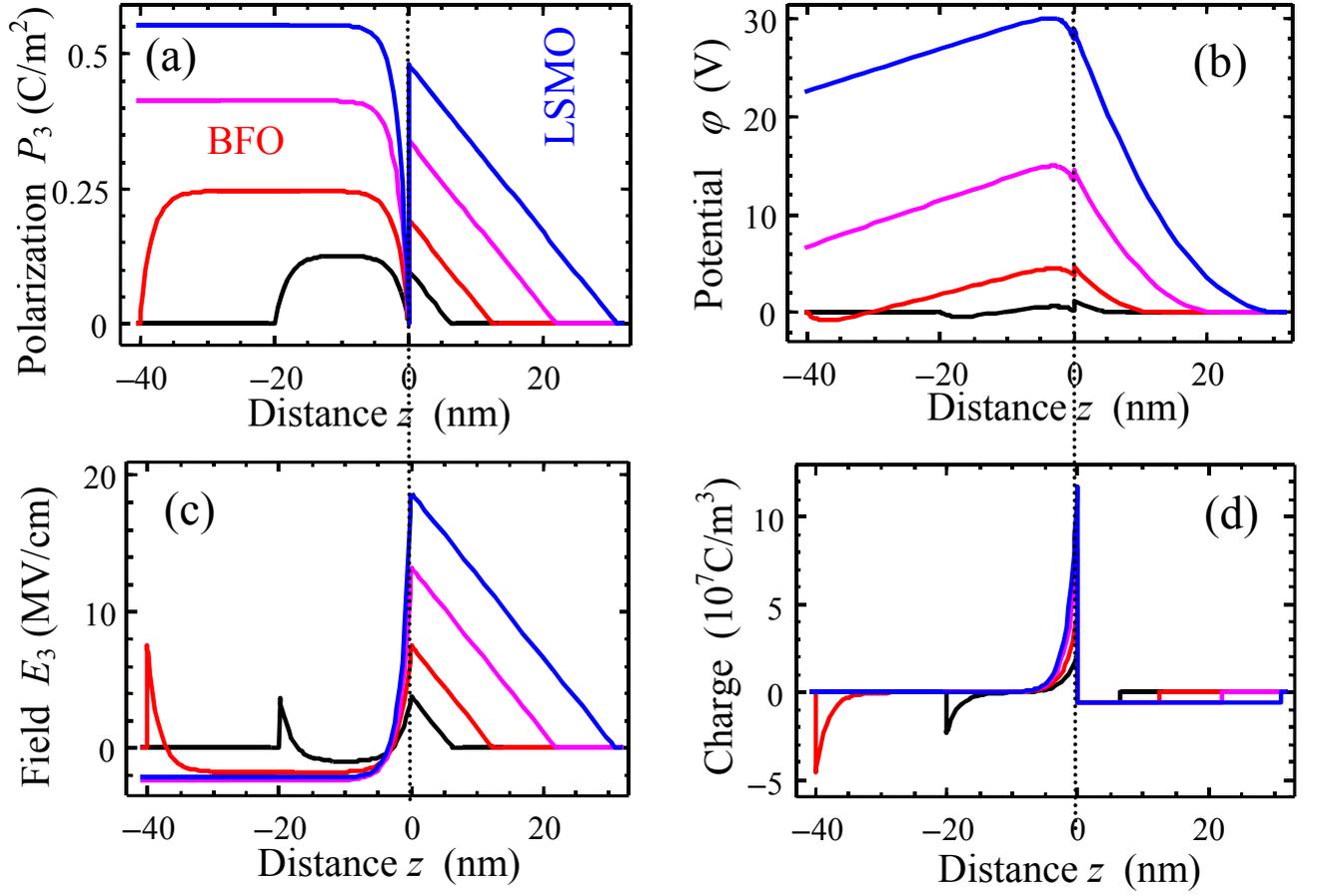

**Fig.5.** (a) Polarization, (b) potential, (c) electric field and (d) electric charge density $z$-distributions for the BFO/LSMO heterostructure. Contact potential difference at $z=0$ is $U_b=1$ V, interface polarization $P_b = 0$ and charge density $\sigma_S = 0$. Carriers concentration in LSMO is $p_S^0 = 10^{26}$ m$^{-3}$. Black, red, violet and blue curves correspond to different values of BFO film thickness $L = 20, 40, 80, 160$ nm with extrapolation lengths $\lambda_i \approx 0$ nm and the gradient coefficient $g = 10^{-8}$ m$^3$/F.

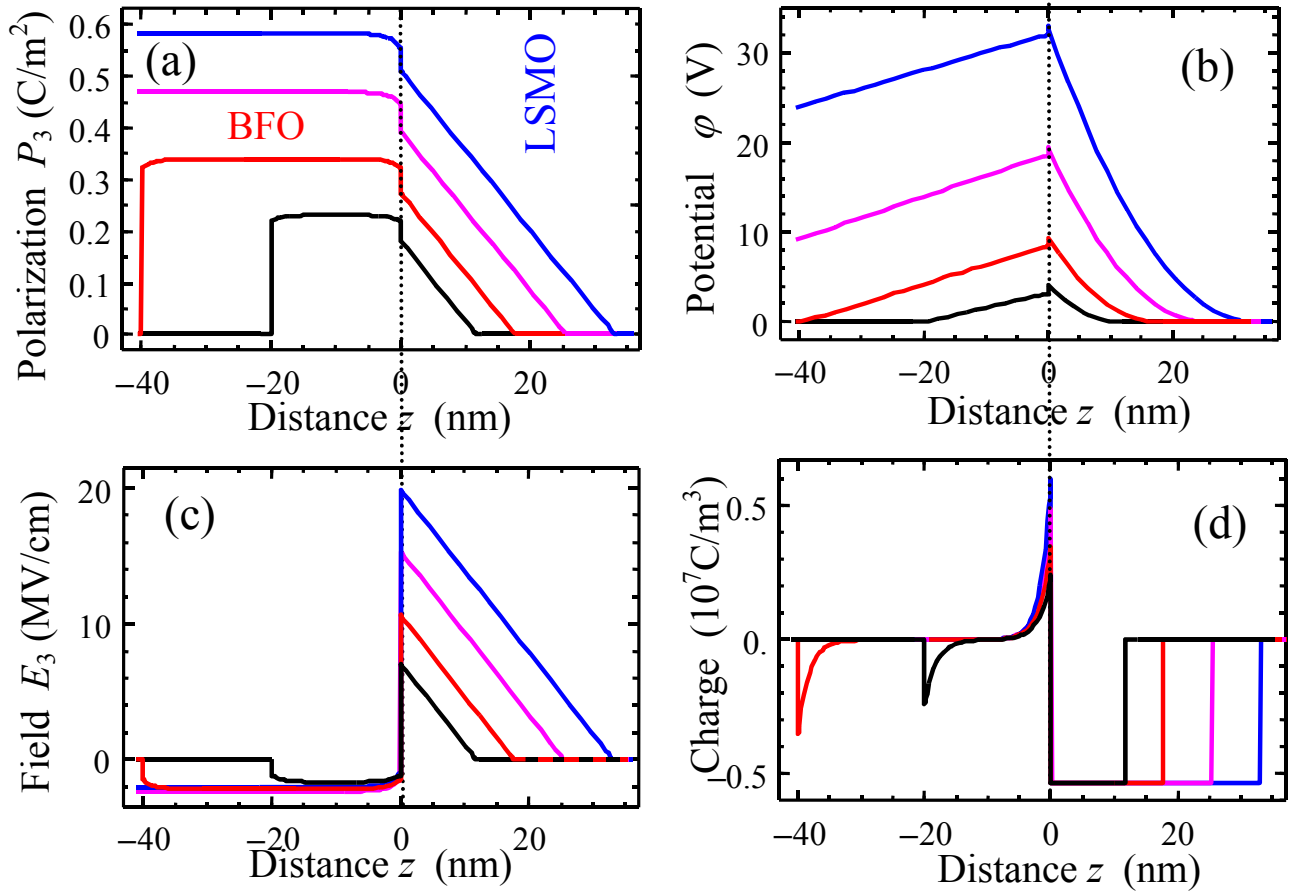

**Fig.6.** (a) Polarization, (b) potential, (c) electric field and (d) electric charge density z-distributions for the BFO/LSMO heterostructure. Contact potential difference at $z$=0 is $U_b$=1 V, interface polarization $P_b = 0$ and charge density $\sigma_S = 0$. Carriers concentration in LSMO is $p_S^0 = 10^{26}$ m$^{-3}$. Black, red, violet and blue curves correspond to different thickness $L$ = 20, 40, 80, 160 nm of BFO film with extrapolation lengths $\lambda_i$=30 nm and the gradient coefficient g = $10^{-8}$ m$^3$/F.

Figs. 7-8 illustrate z-distributions of the electric polarization, potential, field and bulk charge density in the heterostructure "incipient SrTiO$_3$ film – half-metal LaSrMnO$_3$" (STO/LSMO) for zero interface charge $\sigma_S$=0. When generating the plots in Figs. 2 we put $\lambda_i$=0. Plots in Figs. 6 correspond to high enough $\lambda_i$ values. The resulting built-in field

$$E_b^f \sim \left( \frac{U_b}{L}\langle f \rangle - \frac{P_b}{\varepsilon_0 \varepsilon_{33}^b}\langle b \rangle \right)$$ (see Eq.(11b)) induces the electric polarization in the incipient ferroelectric films.

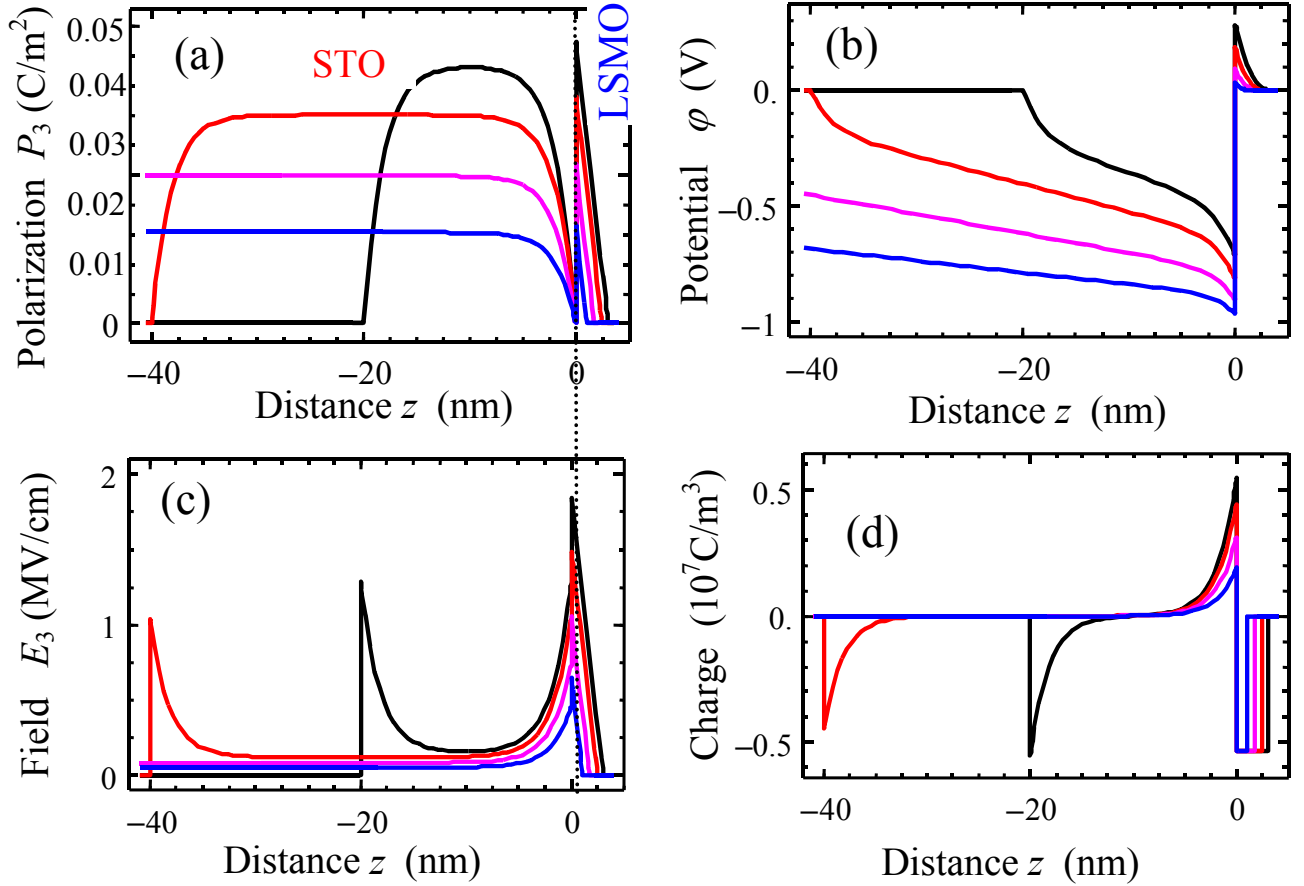

**Fig.7.** (a) Polarization, (b) potential, (c) electric field and (d) electric charge density z-distributions for the STO/LSMO heterostructure. Contact potential difference at $z=0$ is $U_b=1$ V, interface polarization $P_b = 0$ and charge density $\sigma_S = 0$. Carriers concentration in LSMO is $p_S^0 = 10^{26}$ m$^{-3}$. Black, red, violet and blue curves correspond to different thickness $L = 20, 40, 80, 160$ nm of STO film with extrapolation lengths $\lambda_i \approx 0$ nm and the gradient coefficient g = $10^{-8}$ m$^3$/F.

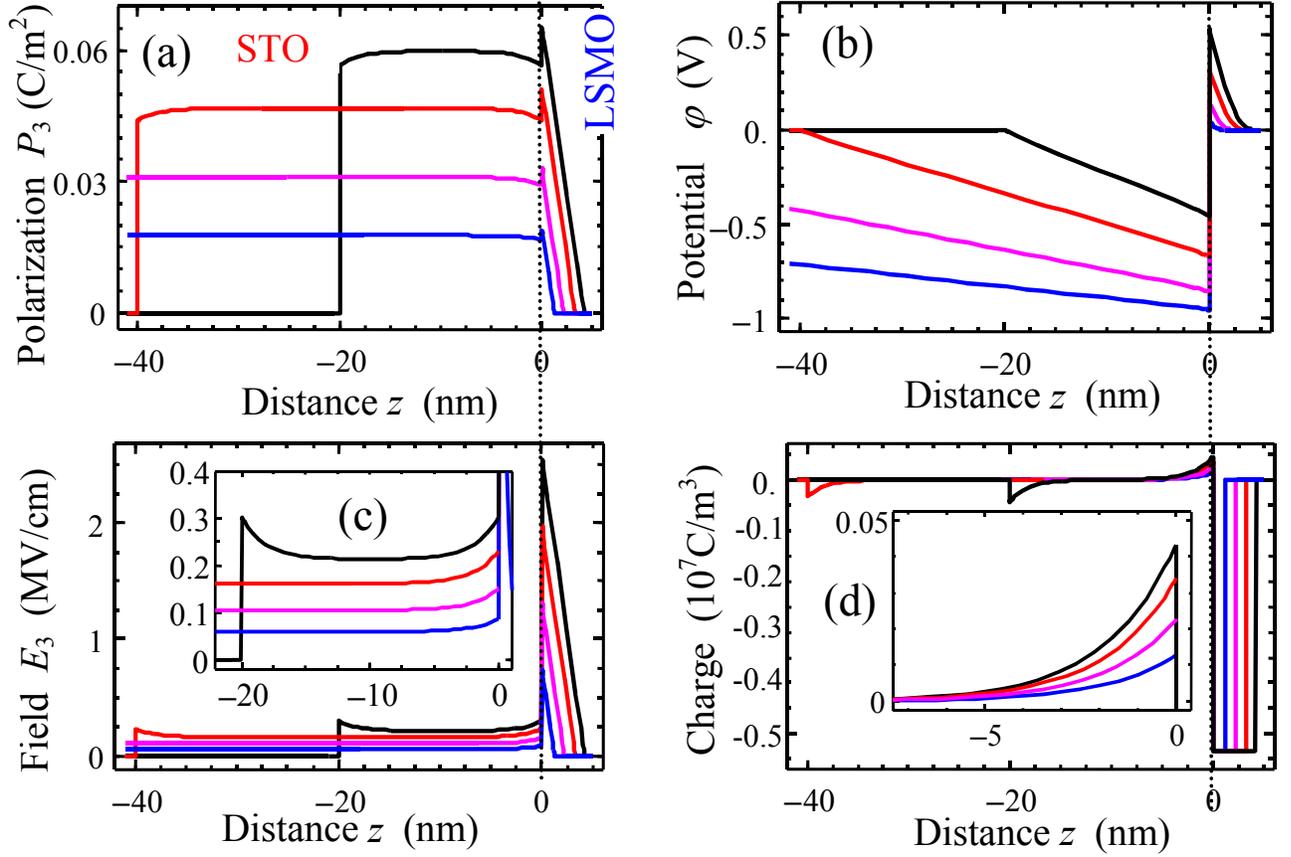

**Fig.8.** (a) Polarization, (b) potential, (c) electric field and (d) electric charge density z-distributions for the STO/LSMO heterostructure. Contact potential difference at $z=0$ is $U_b=1$ V, interface polarization $P_b = 0$ and charge density $\sigma_S = 0$. Carriers concentration in LSMO is $p_S^0 = 10^{26}$ m$^{-3}$. Black, red, violet and blue curves correspond to different thickness $L =$ 20, 40, 80, 160 nm of STO film with extrapolation lengths $\lambda_i=30$ nm and the gradient coefficient g = $10^{-8}$ m$^3$/F.

The distributions shown in Figs. 7-8 correspond to the stable ground state of the heterostructure incipient ferroelectric STO/LSMO without interface charge ($\sigma_S = 0$), i.e. when the free carriers are abundant in LSMO and there is no need in the screening interface charge. The characteristic feature of the interface charge absence is the thin depletion layer $W_S$ (several nm for LSMO) that is occupied by the main-type carriers. The opposite case of depletion layers created by the minor-type carriers, which can appear during the polarization reversal in the proper ferroelectric film, will be considered in the next section.

### 3.3. Calculations of the metastable states for typical heterostructures (1D-case)

As it was mentioned in Sections 2-3, when the free carriers are abundant there is no need in the screening interface charge (i.e. for thin depletion layer created by the main-type carriers and thick layer created by the minor type carries). In the opposite case of depletion layers created only by the minor-type carriers (i.e. without interface charge states located at $z = 0$) the penetration depth $W_S$ is higher (up to tens of nanometers as shown by the bottom curves in Figs. 9) and corresponding screening of the spontaneous polarization appeared weaker (compare bottom curves in Fig. 9a for metastable polarization with the top curves for the stable ground state).

Fig. 10 shows the hysteresis loops of the average polarization in the BFO film and corresponding field penetration depth in LSMO under the absence of the interface charge $\sigma_S$ and two different values of the major-type carriers in LSMO. The loops asymmetry increases with the increase of the carriers concentration [compare Figs.10a,c with Figs.10b,d]. The asymmetry of the loops, both horizontal imprint and vertical shift, are caused by the charge effects provided by the major- type carriers for positive biases $(U_e + U_b) > 0$ and minor-type carriers for negative biases $(U_e + U_b) < 0$ respectively resulting in the appearance of the build-in field.

The weak screening causes strong electric fields, which resulting into the suppression of polarization inside the ferroelectric film and increase of the system's free energy, since depolarization field energy is always positive. As a result, the strong field effect may lead to the bend bending at $z = 0$ and appearance of interface charge states.

Fig. 11 shows the influence of the interface charge on the thickness dependence of the stable and metastable (if any) states of the average polarization $\langle P_3 \rangle$ in ferroelectric BFO film.

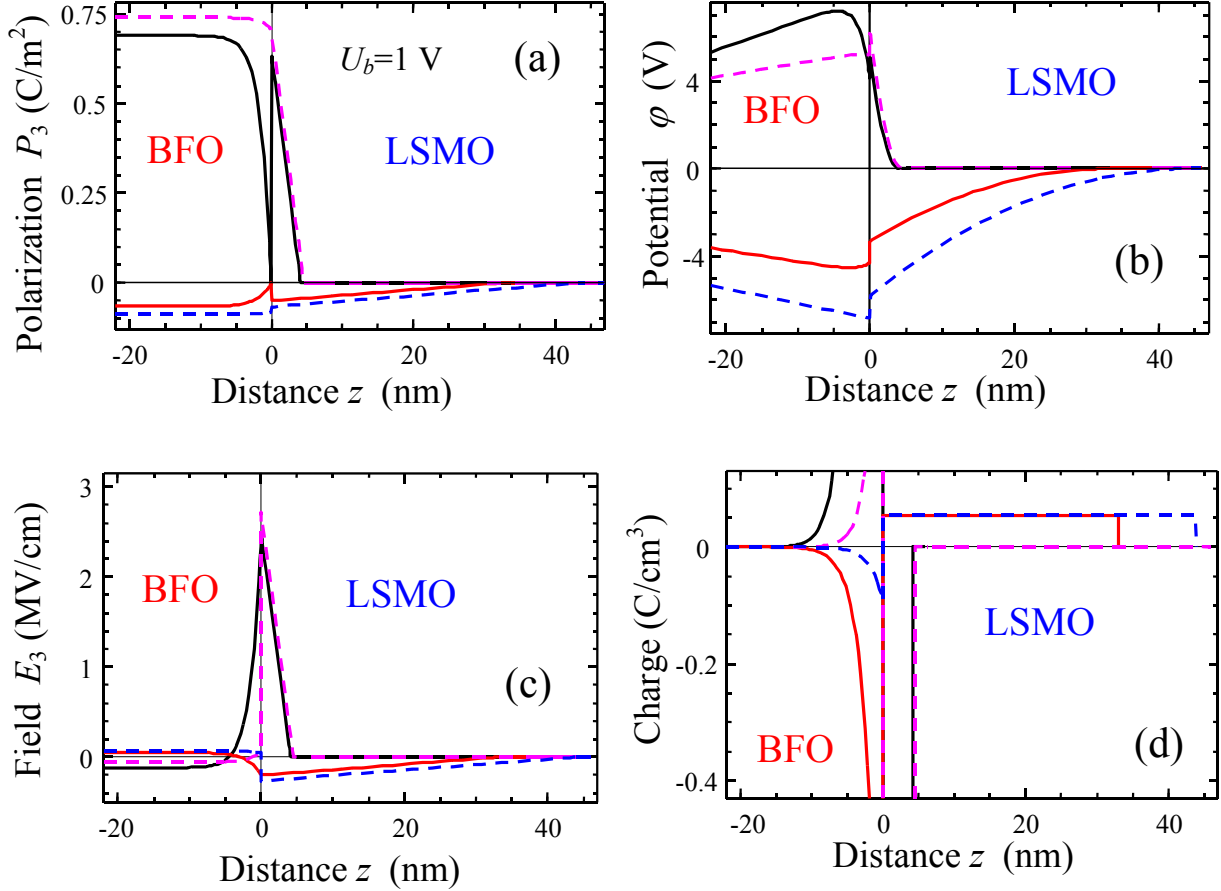

**Fig. 9.** (a) Polarization, (b) potential, (c) electric field and (d) bulk charge density z-distributions near the BFO/LSMO interface at zero external bias $U_e = 0$. Contact potential difference at $z=0$ is $U_b=1$ V, interface polarization $P_b = 0$ and charge density $\sigma_S = 0$. BFO film thickness $L = 100$ nm, gradient coefficient $g = 10^{-8}$ m$^3$/F, $\lambda_i \approx 0$ nm (solid curves) and $\lambda_i = 30$ nm (dashed curves). Carriers concentration in LSMO is $p_S^0 = 10^{27}$ m$^{-3}$ (upper curves) and $n_S^0 = 10^{25}$ m$^{-3}$ (bottom curves). Upper curves (positive $P_3$) correspond to the stable ground states, bottom ones (negative $P_3$) correspond to the reversed polarization (metastable states).

It is seen from Fig. 11 that the interface charges $\sigma_S$ of appropriate sign increases the average polarization and smear the size-induced phase transition point for the stable states (compare upper curves 1-5). Also the interface charges lead to the strong asymmetry of the average polarization values in the stable "up" and metastable "down" states, which exist not for all considered values of $\sigma_S$ (compare up and bottom curves in Figs.11). The interface charges $\sigma_S$ act as the contribution into the built-in field in the right-hand-side of Eqs.(10).

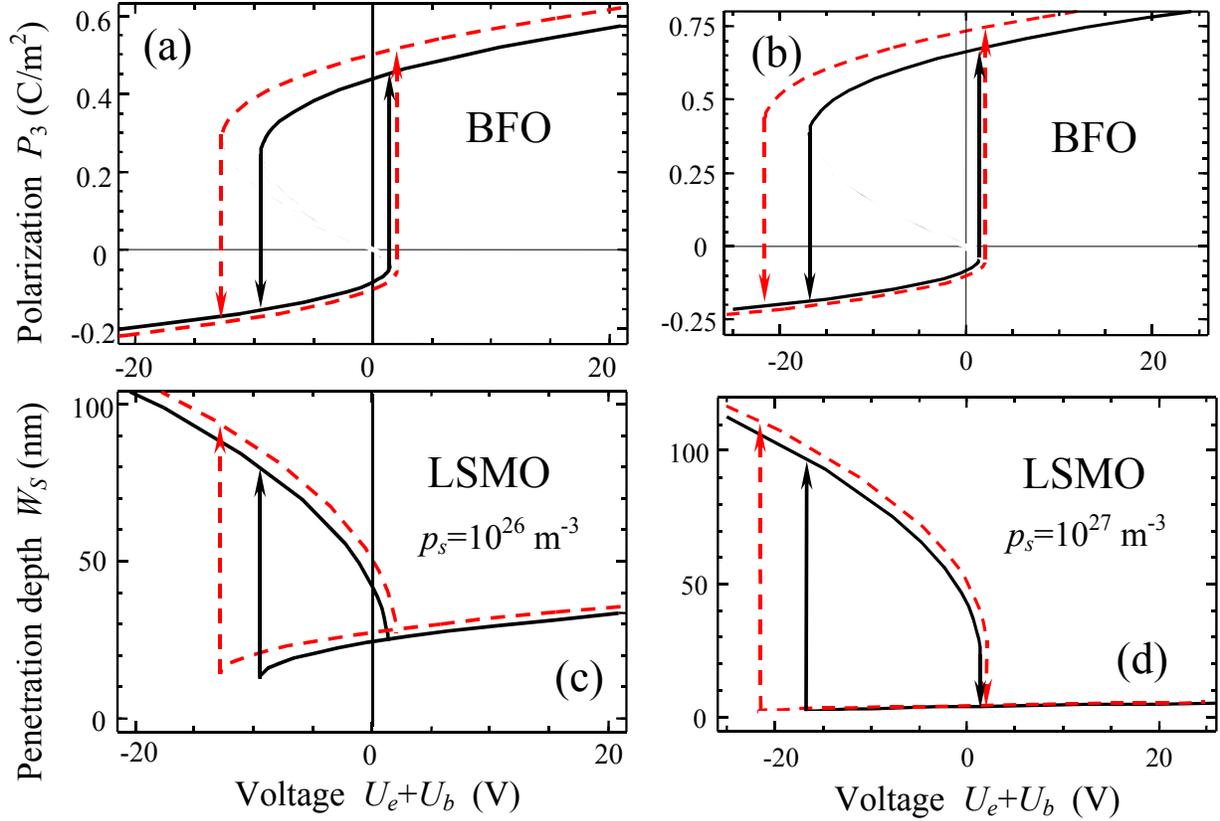

**Fig. 10.** Bias dependence of the average BFO polarization (a, b) and LSMO penetration depth (c, d). BFO film thickness $L = 100$ nm, gradient coefficient g $= 10^{-8}$ m$^3$/F, extrapolation lengths $\lambda_i \approx 0$ nm (solid curves) and $\lambda_i = 30$ nm (dashed curves). Interface polarization $P_b = 0$ and charge density $\sigma_S = 0$. LSMO major-type carrier concentration is $p_S^0 = 10^{26}$ m$^{-3}$ (a, c) and $p_S^0 = 10^{27}$ m$^{-3}$(b, d); and $n_S^0 = 10^{25}$ m$^{-3}$ for minor-type carriers respectively. Gap is absent ($H = 0$).

As expected, the interface charges $\sigma_S$ of appropriate sign provide more effective screening of the spontaneous polarization than the extended space-charge layer. The screening by the interface charges $\sigma_S$ makes polarization more homogeneous, subsequently decreases the depolarization field, which in turn self-consistently decreases the system free energy. So the absolutely stable profiles of reversed polarization shown in Figs.12a by the bottom curves are more energetically preferable than the ones shown by the bottom curves in Figs.9a for zero interface charges ($\sigma_S = 0$). Thus the interface charge may originate in the case of the weak polarization screening from the semiconductor side (i.e. for thick depletion layer created by the minor-type carriers).

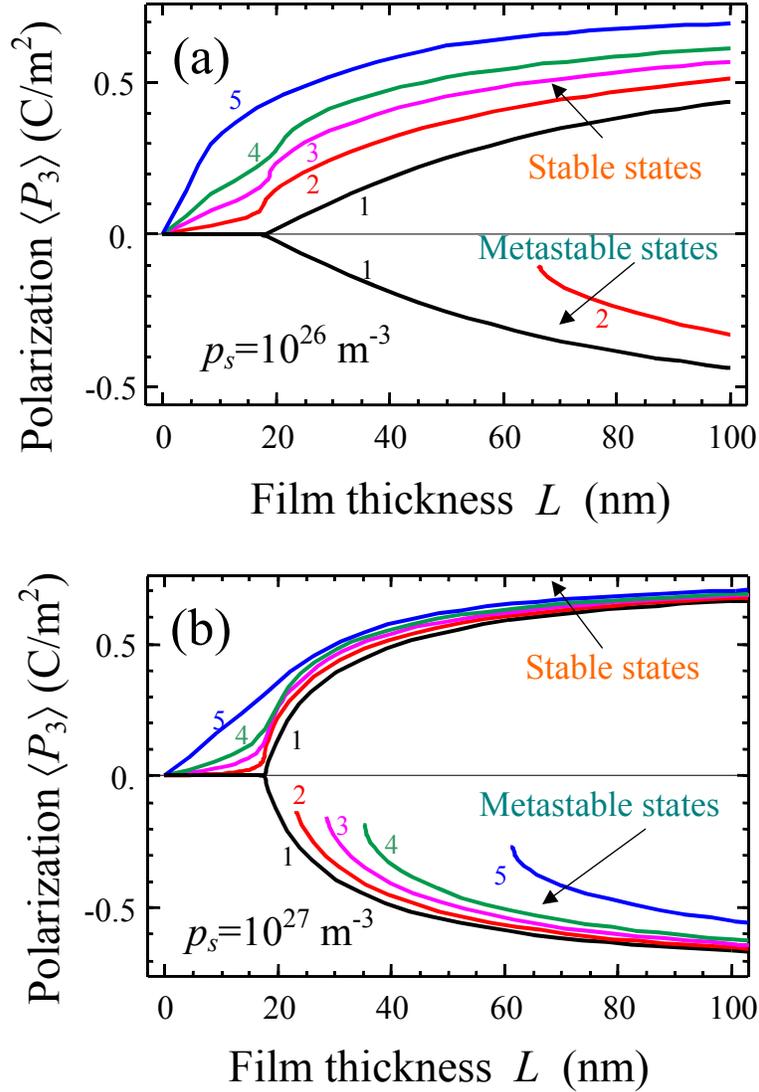

**Fig. 11.** The average BFO polarization thickness dependence for different interface charge density $\sigma_S$ = 0, -0.1, -0.2, -0.3, -0.6 C/m² (curves 1, 2, 3, 4, 5 respectively). Extrapolation lengths $\lambda_i$=0 nm, carriers concentration in semiconductor is $n_S^0 = p_S^0 = 10^{26}$ m$^{-3}$ (a) and $n_S^0 = p_S^0 = 10^{27}$ m$^{-3}$ (b). Other parameters are the same as in Figs.9. Upper curves are the stable polarization states. Metastable states corresponding to negative polarization values are shown by the bottom curves (if any exist for definite $\sigma_S$).

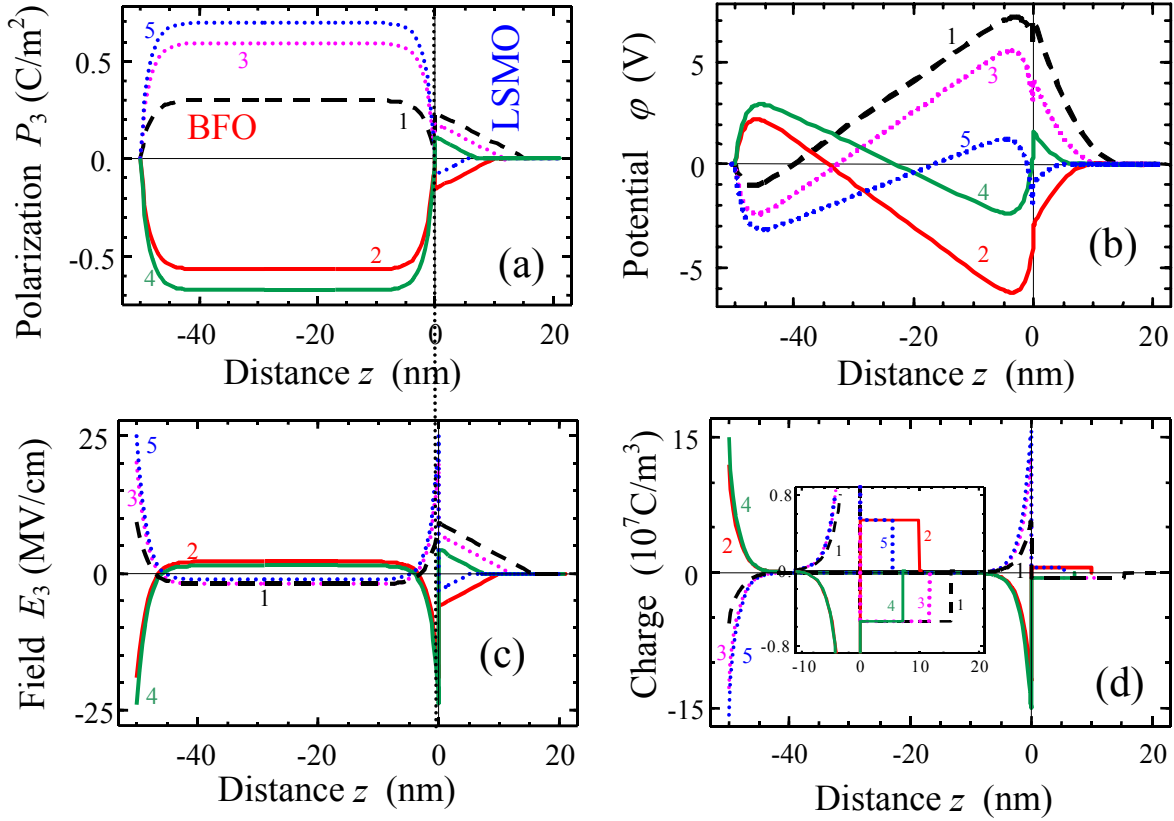

**Fig.12.** (a) Polarization, (b) potential, (c) electric field and electric charge density (d) z-distributions for the STO/LSMO heterostructure at zero external bias $U_e = 0$. Contact potential difference at $z=0$ is $U_b=1$ V, interface polarization $P_b = 0$ and different interface charge density $\sigma_S = 0, +0.35, -0.35, +0.75, -0.75$ C/m$^2$ (curves 1, 2, 3, 4, 5 respectively). Carriers concentration in LSMO is $p_S^0 = 10^{26}$ m$^{-3}$. BFO film thickness $L = 50$ nm, extrapolation lengths $\lambda_i \approx 0$ nm and the gradient coefficient g = $10^{-8}$ m$^3$/F.

## 4. Effect of the incomplete screening and interface charge on the local polarization reversal and charge transport

In the initial ground state the sluggish surface charges $\sigma_f$ completely screen the electric displacement outside the film. In contrast to the ground states considered in the Section 3, the recharging of surface charges $\sigma_f$ should appear during the local polarization reversal. The ultra-thin dielectric gap of thickness $H$ models the separation between the tip electrode and the sluggish charges inside a contamination (or dead) layer.

The equilibrium domain structure is almost cylindrical for the case of local polarization reversal caused by the localized potential $U_e(x,y)$ applied to the SPM-tip in thin ferroelectric film [Fig. 1c]. As a sequence the spontaneous polarization distribution, the potential and the interface charges density variations can be expressed in $\{x,y\}$ - Fourier representation. In the Fourier $k_{1,2}$-domain Eqs.(2, 3) immediately split into the two systems of differential equations. The first system corresponds to the smooth components $\{\varphi(z), P_3(z)\}$ was solved in the previous section. The second system for the modulating components is listed in Appendix B.

Approximate analytical solution (6) may be used in particular case of the disk-like tip apex of radius $R \gg L$, i.e. until $\Delta_\perp U_e(x,y) \approx 0$ in the region of polarization reversal.

It was shown earlier [35] that the transverse polarization gradient could be neglected for the case of the strong inequality $\sqrt{g/2|\alpha|} \ll R$ valid for all considered ferroelectrics at room temperature. Thus polarization $P_3(x,y,z) \sim f(z,L)$ averaged over the film thickness can be determined from the Eqs.(10).

Using the coercive volume conception of polarization reversal formulated in Ref.[35], the domain lateral sizes $\{x,y\}$ can be estimated from the equation:

$$\frac{\varepsilon_{33}^g \langle f \rangle}{\varepsilon_{33}^g L + \varepsilon_{33}^b H} U_e(x,y) = E_c^\pm(L,H). \tag{15}$$

The intrinsic coercive fields are given by Eq.(11b).

For a typical tip potential distribution $U_e(r) \approx \dfrac{Ud}{\sqrt{r^2+d^2}}$ ($d$ is the effective tip size) the domain radius $r = \sqrt{x^2+y^2}$ depends on applied bias $U$ as

$$r(U) = d\sqrt{U^2 \left(\frac{\varepsilon_{33}^g L + \varepsilon_{33}^b H}{\varepsilon_{33}^g \langle f \rangle} E_c^\pm \right)^{-2} - 1}.$$

In general case the current density inside ferroelectric film consists of the conductivity, diffusion and displacement, tunneling, Schottky and Frenkel-Poole emission currents [36]. During the local polarization reversal the thickness and electric charge of the space charge layer changes, it particular the positive charge can be substituted by the negative one or vise versa depending on the polarization direction. Such transformations should be accompanied by the peaks of displacement currents, which shape and amplitude depend on the domain shape and sizes. However, for nanoscale contact a displacement current is significantly smaller and faster then a constant leakage current, and hence is ignored.

Derived analytical expressions also allow estimations of the tunneling current between the tip apex and ferroelectric surface (if any exist). Assuming that the ultra-thin gap $H$ is transparent for the tunneling electrons, tunneling and field emission currents could flow between the tip apex and ferroelectric film surface. In the Fowler-Nordheim transport regime [36, 37] the tunneling current density $J_t \sim \exp\left(-\frac{2}{\hbar}\int_{-L-H}^{0} dz\sqrt{2m^*q|\varphi(z)|}\right)$ is determined by the potential distribution $\varphi(z)$ given by Eq.(6) and corresponding penetration depth in semiconductor, $W_S(U_e, L, H)$, is determined self-consistently from Eqs.(10).

## 5. Summary

Using Landau-Ginzburg-Devonshire approach we have calculated the equilibrium distributions of electric field, polarization and space charge in the ferroelectric-semiconductor heterostructures containing proper or incipient ferroelectric thin films. In particular, it is shown that space charge effects introduce strong size-effect on spontaneous polarization in 20-40 nm epitaxial films of $BiFeO_3$ on $(LaSr)MnO3$, and can induce strong polarization in incipient ferroelectric $SrTiO_3$.

We obtained analytical expressions for the cylindrical domain sizes appeared in ferroelectric film under the local polarization reversal, which is caused by the electric field induced by the nanosized tip of the SPM probe. The SPM tip can be separated from the ferroelectric surface covered with sluggish screening charges by the ultra-thin dielectric layer that models either geometric gap or/and contamination layer resistance.

The intrinsic field effects, which originated at the ferroelectric-semiconductor interface, lead to the surface band bending and result in the formation of depletion/accumulation space-charge layer near the semiconductor surface. We calculated how the build-in fields smear the size-induced phase transition, induce polarization in the incipient ferroelectric films and lead to the polarization hysteresis loops vertical and horizontal asymmetry in ferroelectric films.

**Acknowledgements**

Authors are grateful to Prof. E. Tsymbal for valuable critical remarks. Research is sponsored by Ministry of Science and Education of Ukraine and National Science Foundation (Materials World Network, DMR-0908718). SVK and AB acknowledge the DOE SISGR program.

**Appendix A.**

Allowing for Eq.(7), polarization distribution $P_3(z)$ should be found from the Euler-Lagrange boundary problem (2) as:

$$\begin{cases} \left(\alpha + \dfrac{1}{\varepsilon_0 \varepsilon_{33}^b}\right) P_3 + \beta P_3^3 - g \dfrac{d^2}{dz^2} P_3 = \dfrac{\rho_S^0 W_S - \sigma_S}{\varepsilon_0 \varepsilon_{33}^b}, \\ \left(P_3 + \lambda_1 \dfrac{dP_3}{dz}\right)\bigg|_{z=0} = -P_b, \quad \left(P_3 - \lambda_2 \dfrac{dP_3}{dz}\right)\bigg|_{z=-L} = 0. \end{cases} \quad (A.1)$$

Let us look for the solution of the problem (A.1) in the form $P_3(z) = \langle P_3 \rangle + p(z)$, where the average polarization $\langle P_3 \rangle \equiv \dfrac{1}{L}\int_{-L}^{0} P_3(\tilde{z}) d\tilde{z}$ is introduced. The variation $p$ average value is zero: $\langle p \rangle \equiv 0$. So, the problem (A.1) acquires the form:

$$\begin{cases} \left(\alpha + 3\beta\langle P_3 \rangle^2 + \dfrac{1}{\varepsilon_0 \varepsilon_{33}^b}\right) p + 3\beta\langle P_3 \rangle p^2 + \beta p^3 - g \dfrac{d^2 p}{dz^2} = \dfrac{\rho_S^0 W_S - \sigma_S - \langle P_3 \rangle}{\varepsilon_0 \varepsilon_{33}^b} - \left(\alpha\langle P_3 \rangle + \beta\langle P_3 \rangle^3\right), \\ \left(p + \lambda_1 \dfrac{dp}{dz}\right)\bigg|_{z=0} = -\langle P_3 \rangle - P_b, \quad \left(p - \lambda_2 \dfrac{dp}{dz}\right)\bigg|_{z=-L} = -\langle P_3 \rangle. \end{cases}$$

(A.2)

Since always $\alpha + \dfrac{1}{\varepsilon_0 \varepsilon_{33}^b} \gg 0$ (as well as $\alpha + 3\beta\langle P_3 \rangle^2 + \dfrac{1}{\varepsilon_0 \varepsilon_{33}^b} \gg 0$) for both proper and incipient ferroelectrics, Eq.(A.2) can be linearized with respect to the deviation $p$ and then solved by standard methods.

After elementary transformations, polarization distribution acquires the form:

$$P_3(z) = \dfrac{\varepsilon_S - 1}{\varepsilon_S} q \begin{cases} p_S^0(z - W_{Sp}) \cdot \theta(W_{Sp} - z), & z \geq 0, \text{ depletion of n-type carriers} \\ -n_S^0(z - W_{Sn}) \cdot \theta(W_{Sn} - z), & z \geq 0, \text{ depletion of p-type carriers} \end{cases} \quad (A.3)$$

$$P_3(z) = \frac{2\varepsilon_0\varepsilon_{33}^b\beta\langle P_3\rangle^3 + \rho_S^0 W_S - \sigma_S}{\varepsilon_0\varepsilon_{33}^b(\alpha + 3\beta\langle P_3\rangle^2) + 1} f(z,L) - P_b \cdot b(z,L), \quad -L < z \leq 0. \quad (A.4)$$

The space distribution is governed by the functions $f$ and $b$:

$$f(z,L) = 1 - \xi\frac{\lambda_2\cosh((L+z)/\xi) + \lambda_1\cosh(z/\xi) + \xi(\sinh((L+z)/\xi) - \sinh(z/\xi))}{(\xi^2 + \lambda_1\lambda_2)\sinh(L/\xi) + \xi(\lambda_1 + \lambda_2)\cosh(L/\xi)}, \quad (A.5a)$$

$$b(z,L) = \frac{\xi\lambda_2\cosh((L+z)/\xi) + \xi^2\sinh((L+z)/\xi)}{(\xi^2 + \lambda_1\lambda_2)\sinh(L/\xi) + \xi(\lambda_1 + \lambda_2)\cosh(L/\xi)}. \quad (A.5b)$$

Characteristic length $\xi = \sqrt{\dfrac{\varepsilon_0\varepsilon_{33}^b g}{\varepsilon_0\varepsilon_{33}^b(\alpha + 3\beta\langle P_3\rangle^2) + 1}} \approx \sqrt{\varepsilon_0\varepsilon_{33}^b g}$.

Then the average polarization $\langle P_3\rangle$ and depths $W_{Sn,p}$ should be determined self-consistently from the spatial averaging of Eq.(8b),

$$\langle P_3\rangle = \frac{2\varepsilon_0\varepsilon_{33}^b\beta\langle P_3\rangle^3 + \rho_S^0 W_S - \sigma_S}{\varepsilon_0\varepsilon_{33}^b(\alpha + 3\beta\langle P_3\rangle^2) + 1}\langle f\rangle - P_b\cdot\langle b\rangle, \text{ and boundary conditions (4b,c).}$$

For particular case $H = 0$ this gives the system of coupled algebraic equations:

$$\begin{cases} -\dfrac{\rho_S^0 W_S^2}{2\varepsilon_0\varepsilon_S} + L\left(\dfrac{\rho_S^0 W_S}{\varepsilon_0\varepsilon_{33}^b} - \dfrac{\sigma_S + \langle P_3\rangle}{\varepsilon_0\varepsilon_{33}^b}\right) = U_b, \\ \langle P_3\rangle\left(\alpha + \beta\langle P_3\rangle^2(3 - 2\langle f\rangle) + \dfrac{1}{\varepsilon_0\varepsilon_{33}^b}\right) = \left(\dfrac{\rho_S^0 W_S - \sigma_S}{\varepsilon_0\varepsilon_{33}^b}\right)\langle f\rangle - P_b\cdot\langle b\rangle\left(\alpha + 3\beta\langle P_3\rangle^2 + \dfrac{1}{\varepsilon_0\varepsilon_{33}^b}\right). \end{cases} \quad (A.6)$$

Where $\langle f\rangle = 1 - \dfrac{\xi^2(2\xi(\cosh(L/\xi) - 1) + (\lambda_1 + \lambda_2)\sinh(L/\xi))}{L(\xi(\lambda_1 + \lambda_2)\cosh(L/\xi) + (\xi^2 + \lambda_1\lambda_2)\sinh(L/\xi))} \approx 1 - \dfrac{\xi^2(2\xi + \lambda_1 + \lambda_2)}{L(\xi(\lambda_1 + \lambda_2) + \xi^2 + \lambda_1\lambda_2)}$ and

$$\langle b\rangle = \frac{\xi^2\lambda_2\sinh(L/\xi) - \xi^3(1 - \cosh(L/\xi))}{L((\xi^2 + \lambda_1\lambda_2)\sinh(L/\xi) + \xi(\lambda_1 + \lambda_2)\cosh(L/\xi))} \approx \frac{\xi^2(\xi + \lambda_2)}{L(\xi^2 + \lambda_1\lambda_2 + \xi(\lambda_1 + \lambda_2))} = \frac{\xi^2}{L(\xi + \lambda_1)}.$$

The system (A.6) reduces to the relations

$$\begin{cases} \langle P_3\rangle = -\sigma_S + \rho_S^0 W_S + \dfrac{\varepsilon_{33}^b\rho_S^0}{2L\varepsilon_S}W_S^2 - \varepsilon_0\varepsilon_{33}^b\dfrac{U_b}{L}, \\ \left(\alpha + \dfrac{1}{\varepsilon_0\varepsilon_{33}^b}\right)\langle P_3\rangle + \beta\langle P_3\rangle^3(3 - 2\langle f\rangle) = \left(\dfrac{\rho_S^0 W_S - \sigma_S}{\varepsilon_0\varepsilon_{33}^b}\right)\langle f\rangle - \dfrac{P_b\langle b\rangle}{\varepsilon_0\varepsilon_{33}^b}. \end{cases} \quad (A.7)$$

Since $\langle P_3 \rangle = -\sigma_S + \rho_S^0 W_S + \dfrac{\varepsilon_{33}^b \rho_S^0}{2L\varepsilon_S} W_S^2 - \varepsilon_0 \varepsilon_{33}^b \dfrac{U_b}{L}$ and $\left(\langle P_3 \rangle + \sigma_S + \varepsilon_0 \varepsilon_{33}^b \dfrac{U_b}{L}\right)\dfrac{1}{\rho_S^0} > 0$, the second of Eqs.(14) reduces to six order algebraic equation for the built-in field determination.

The polarization contribution into the relative atomic displacement $u_3$ can be estimated as $u_3(z) \approx \dfrac{V_S}{Q_S^B} P_3(z)$ at $z > 0$ and $u_3(z) \approx \dfrac{V_{FE}}{Q_{IF}^B} P_3(z)$ at $-L < z < 0$. Here $V_j$ is the volume of the corresponding unit cell, $Q^B$ is the Born effective charge of the lightest atom "B". For perovskites considered hereinafter $V_{FE,S} \approx 6.4 \cdot 10^{-29}$ m$^3$

**Appendix B.**

The potential applied to the SPM-tip is highly-localized, i.e. $U(x,y) = \int_{-\infty}^{\infty} dk_1 \int_{-\infty}^{\infty} dk_2 u(\mathbf{k}) \exp(-ik_1 x - ik_2 y)$. As a sequence the spontaneous polarization distribution can be approximated as $P_S(x,y,z) = P_3(z) + \int_{-\infty}^{\infty} dk_1 \int_{-\infty}^{\infty} dk_2 p_S(\mathbf{k},z) \exp(-ik_1 x - ik_2 y)$, the potential $\varphi_f(x,y,z) = \varphi(z) + \int_{-\infty}^{\infty} dk_1 \int_{-\infty}^{\infty} dk_2 \widetilde{\varphi}(\mathbf{k},z) \exp(-ik_1 x - ik_2 y)$ and the interface charges density variation $\delta\sigma_S(x,y) = \int_{-\infty}^{\infty} dk_1 \int_{-\infty}^{\infty} dk_2 \widetilde{\sigma}_S(\mathbf{k}) \exp(-ik_1 x - ik_2 y)$ can be expressed in Fourier representation (see).

In the Fourier $k_{1,2}$-domain Eqs.(3) immediately split into the two systems of differential equations. The first system corresponds to the smooth components $\{\varphi(z), P_3(z)\}$ was solved in the previous section. The second system for the modulating components is:

$$\begin{aligned} &\dfrac{d^2\widetilde{\varphi}}{dz^2} - k^2\widetilde{\varphi} = 0, \qquad -H-L < z < -L, \\ &\varepsilon_{33}^b \dfrac{d^2\widetilde{\varphi}}{dz^2} - \varepsilon_{11} k^2 \widetilde{\varphi} = \dfrac{1}{\varepsilon_0} \dfrac{dp_S(\mathbf{k},z)}{dz}, \qquad -L < z < 0, \\ &\dfrac{d^2\widetilde{\varphi}}{dz^2} - k^2\widetilde{\varphi} = 0, \quad z > 0. \end{aligned} \qquad (B.1)$$

Where $k^2 = k_1^2 + k_2^2$. Rewritten for the modulating components, the boundary conditions (4) acquire the form:

$$\begin{aligned}
&\widetilde{\varphi}(k,-L-H) = u(k), \\
&\widetilde{\varphi}(k,-L+0) - \widetilde{\varphi}(k,-L-0) \approx 0, \\
&\varepsilon_0\left(\varepsilon_{33}^b \frac{d\widetilde{\varphi}(k,-L+0)}{dz} - \varepsilon_{33}^g \frac{d\widetilde{\varphi}(k,-L-0)}{dz}\right) - p_S(k,-L) = \widetilde{\sigma}_f(k), \\
&\varepsilon_0\left(\varepsilon_{33}^b \frac{d\widetilde{\varphi}(k,+0)}{dz} - \varepsilon_S \frac{d\widetilde{\varphi}(k,-0)}{dz}\right) - p_S(k,0) = \widetilde{\sigma}_S(k), \quad \widetilde{\varphi}(k,+0) - \widetilde{\varphi}(k,-0) \approx 0, \\
&\widetilde{\varphi}(k, z \to \infty) = 0.
\end{aligned} \quad (B.2)$$

Where the gap dielectric permittivity $\varepsilon_{33}^g$ is introduced.

LGD-equation (2) for determination of the modulation $p_S(\mathbf{k},z)$ acquires the form:

$$\left(\alpha + g k^2 + 3\beta P_3^2(z) - g\frac{d^2}{dz^2}\right) p_S(\mathbf{k}, z) = -\frac{d}{dz}\widetilde{\varphi}(\mathbf{k}, z) - \beta Q[p_S(\mathbf{k}, z)],$$

$$Q[p_S(\mathbf{k},z)] = \int_{-\infty}^{\infty} d\mathbf{k}' p_S(\mathbf{k}') \int_{-\infty}^{\infty} d\mathbf{k}'' p_S(\mathbf{k}-\mathbf{k}'-\mathbf{k}'') p_S(\mathbf{k}'') + 3P_3(z) \int_{-\infty}^{\infty} d\mathbf{k}' p_S(\mathbf{k}-\mathbf{k}') p_S(\mathbf{k}'). \quad (B.3)$$

$$\left(p_S + \lambda_1 \frac{dp_S}{dz}\right)\bigg|_{z=0} = 0, \quad \left(p_S - \lambda_2 \frac{dp_S}{dz}\right)\bigg|_{z=-L} = 0$$

The value $p_S(\mathbf{k},-L)$ should be determined self-consistently. In the final state the distribution of the surface charge $\widetilde{\sigma}_f(\mathbf{k})$ localized at $z = -L$ and the interface charge $\widetilde{\sigma}_S(\mathbf{k})$ localized at $z = 0$ should provide the most effective screening of the spontaneous polarization outside the film and minimal depolarization field energy.

The expression for the bias-dependent barrier height related with applied bias difference and polarization changes is given by expression

$$\delta\Phi_B(x,y) \approx q\left(\int_{-\infty}^{\infty} dk_1 \int_{-\infty}^{\infty} dk_2 \widetilde{\varphi}(\mathbf{k},-L)\exp(-ik_1 x - ik_2 y) - U(x,y)\right). \quad (B.4)$$